\documentclass[twocolumn]{aastex631}

\usepackage[final]{paperclaims}  %
\usepackage{amsmath,amssymb}
\usepackage{graphicx}
\usepackage{hyperref}
\usepackage{natbib}
\usepackage{booktabs}
\usepackage{xspace}

\newcommand{\omh}{\ensuremath{\Omega_m h^2}\xspace}
\newcommand{\obh}{\ensuremath{\Omega_b h^2}\xspace}
\newcommand{\och}{\ensuremath{\Omega_c h^2}\xspace}
\newcommand{\thetastar}{\ensuremath{\theta_\star}\xspace}
\newcommand{\czero}{\ensuremath{c_0}\xspace}
\newcommand{\cone}{\ensuremath{c_1}\xspace}

\newcommand{\LCDM}{\ensuremath{\Lambda\mathrm{CDM}}\xspace}
\newcommand{\wowa}{\ensuremath{w_0\text{--}w_a}\xspace}

\newcommand{\Omk}{\ensuremath{\Omega_k}\xspace}
\newcommand{\rs}{\ensuremath{r_d}\xspace}
\newcommand{\DV}{\ensuremath{D_V}\xspace}
\newcommand{\DM}{\ensuremath{D_M}\xspace}
\newcommand{\DHubble}{\ensuremath{D_H}\xspace}
\newcommand{\sigres}{\ensuremath{\sigma_{\mathrm{res}}}\xspace}
\newcommand{\MB}{\ensuremath{M_B}\xspace}

\begin{document}

\title{Universal distance modes from DESI BAO and Type Ia supernovae:\\
       what do cosmological rulers actually measure?}

\author{Matias Zaldarriaga}
\affiliation{Institute for Advanced Study, Princeton, NJ 08540, USA}

\begin{abstract}
We use an SVD decomposition of the low-redshift distance measurements from DESI BAO and three Type~Ia supernova compilations to identify the leading linear directions probed by the data and to localize the tension with the \LCDM CMB-anchored predictions. The leading direction $V_0$ — whose data amplitude we denote \czero — is, to high accuracy, a measurement of \omh: the projection of the data on $V_0$ probes essentially this one CMB-derived parameter combination. BAO constrains this parameter more tightly than the CMB itself; the three SN compilations do not. In every extension of \LCDM we consider, the leading measurable direction remains $V_0$, and it is where most of the tension with the CMB resides. In the \wowa extension a second direction $V_1$ becomes measurable and provides an independent test of dynamical dark energy; the data show no significant tension in this direction. The only other beyond-\LCDM extension that opens a genuinely new measurable direction is spatial curvature, and only marginally and only for BAO; both measurable directions then independently prefer positive spatial curvature, though the second direction is poorly constrained.
\end{abstract}


\section{Introduction}
\label{sec:intro}

The Dark Energy Spectroscopic Instrument (DESI) has measured baryon acoustic oscillations (BAO) across $0 < z < 2.3$ with unprecedented precision \citep{DESI_DR2_BAO}. Combined with cosmic microwave background (CMB) data and Type~Ia supernovae (SNe~Ia), the DESI collaboration reports a $2.8$--$3.9\sigma$ preference (depending on SN dataset) for evolving dark energy in the \wowa parameterization \citep{DESI_DR2_cosmo}. What drives this preference? We show that it arises primarily from what could be interpreted as a $\sim 2\sigma$ discrepancy in \omh between the CMB and low-redshift distance data, rather than from any dark energy signal.

\claim{w0wa_is_c0}{The DESI \wowa preference comes entirely from \czero (\omh), not from the dark energy direction.}

There are many ways to search for deviations from \LCDM. Here we take \LCDM as a well-established baseline whose parameters are constrained by the CMB (we use ACT DR6; \citealt{ACT_DR6}) to a small allowed range. The question we ask is: \emph{how large is the space of deviations that low-redshift distance probes can actually detect?}

The answer is simpler than one might expect. The CMB-allowed \LCDM parameter range maps into a low-dimensional subspace of distance predictions. For both BAO and SN measurements, the effective dimensionality of this map is one: a single SVD direction $V_0$ along which the CMB allows essentially all of the model motion. We write \czero for the projection of the data onto $V_0$. The matrix being decomposed has one row per CMB chain sample and one column per whitened observable (Section~\ref{sec:svd_method}); its leading singular value exceeds the next by a factor of $\sim 25$ for BAO. The result depends only on the CMB parameter uncertainties and the low-redshift measurement covariance --- not on the low-redshift data values.

We can then ask the same question for extensions beyond \LCDM: does a given extension (dynamical dark energy, spatial curvature, etc.) open a genuinely new measurable direction in observable space, beyond $V_0$? Again, this is answerable from the theory and covariance alone.

The result is a set of predetermined directions onto which we project the data. Searching for deviations reduces to reading off a few numbers (tensions in units of $\sigma$, one per mode).

\textbf{Preview of results.} All tension concentrates in a single universal direction $V_0$: BAO shows $+2.2\sigma$ in \czero against the CMB prediction, while the three SN compilations are individually mild (DES-Dovekie $-0.8\sigma$, Pantheon+ $-1.1\sigma$, Union3 $-1.7\sigma$) and do not have enough constraining power in \czero to confirm or refute the BAO signal. The independent dark-energy direction $V_1$, which becomes measurable in the \wowa extension, shows only $+1.2\sigma$ for BAO --- consistent with a cosmological constant. The only other extension that opens a measurable new direction is spatial curvature, and only for BAO.

Among the other extensions we considered, the phenomenological lensing rescaling $A_\mathrm{lens}$ --- a multiplicative factor on the smoothing of CMB acoustic peaks by gravitational lensing, with $A_\mathrm{lens} = 1$ in \LCDM \citep{Calabrese2008} --- is notable: it reduces the tension by shifting the predicted \czero mean and increasing its variance, reminding us of the fact that the CMB constraint on \omh has a lensing component. The physics is the same as the well-known effect of shifting the reionization optical depth $\tau$: in the CMB, $A_s e^{-2\tau}$ controls the lensing amplitude, so a high-$\tau$ universe and a high-$A_\mathrm{lens}$ universe are nearly degenerate \citep{Sailer2026}. In their analysis \citet{Sailer2026} make this explicit: dropping Planck's low-$\ell$ polarization and letting the rest of the CMB$+$BAO data float $\tau$ within \LCDM yields $\tau = 0.090 \pm 0.012$, $\sim 3$--$5\sigma$ above the Planck low-$\ell$ value. Since essentially all current cosmological analyses simply adopt the Planck $\tau$ measurement, a systematic in that single number would propagate identically into essentially every CMB-anchored cosmological analysis.

\textbf{Relation to previous work.} Several recent analyses have examined the DESI results from angles closely related to ours. \citet{Efstathiou2025_BAO} rotates the BAO distance measurements into directions parallel and perpendicular to the Planck \LCDM degeneracy. The parallel direction effectively measures $\omega_m$, showing a ${\sim}2.1\sigma$ tension with Planck---in quantitative agreement with our $+2.2\sigma$ BAO \czero tension. The perpendicular direction and the equation of state at the pivot redshift, $w(z_{\rm piv} = 0.5) = -0.996 \pm 0.046$, show no departure from \LCDM, consistent with our finding that the dark energy direction \cone is null ($+1.2\sigma$). \citet{Efstathiou2025_SN} has also identified potential photometric systematics in the DES~Y5 supernova compilation, finding a ${\sim}0.04$~mag offset between low and high redshift subsamples; we use the recalibrated DES-Dovekie sample \citep{Popovic2025} which partially addresses these concerns. Our DES-Dovekie \czero tension is indeed the mildest of the three SN datasets ($-0.8\sigma$) but none is significant.

\citet{Ye2025} independently confirm the ${\sim}2\sigma$ BAO--CMB tension using Bayesian suspiciousness, and find that the phantom crossing redshift $z_{\rm pc} = 0.45 \pm 0.04$ coincides with the best-constrained scale---matching our BAO \cone pivot $z = 0.46$. The pivot redshift approach was pioneered by \citet{CortesLiddle2024} in the context of DESI~DR1, who identified what they term the ``PhantomX coincidence'': that $w$ crosses $-1$ precisely at the epoch of observation. Our SVD analysis provides an explanation for this coincidence: the apparent $w \neq -1$ at the \czero pivot is \omh tension dressed in dark energy clothing, while the genuinely independent dark energy direction (\cone) is consistent with a cosmological constant at its own pivot. Note that \czero, although where the tension lives, is a near-perfect proxy for \omh --- a parameter that the CMB also constrains tightly along the same combination --- so the tension can in principle be resolved by any mechanism that shifts the CMB inference of \omh (e.g., $\tau$, $A_\mathrm{lens}$; Section~\ref{sec:extensions}). The dark-energy direction \cone, by construction, cannot be moved by such mechanisms --- which makes its null tension at $+1.2\sigma$ a genuine constraint on dynamical dark energy.

Our $V_0$ direction is the empirical, multi-probe counterpart of the analytically constructed Model-independent Expansion Discrepancy Index (MEDI) of \citet{Weiner2026_MEDI}: both single out the same parameter combination from very different starting points. The MEDI identifies $\omega_m r_d^2$ as the combination that acoustic-scale measurements constrain; within \LCDM, where $r_d$ is nearly fixed by early-universe physics, this reduces to \omh, which is why \czero is almost purely an \omh mode. The physical origin of the \omh dominance in BAO was recognized by \citet{Eisenstein2004} and \citet{Knox2006}. Appendix~\ref{app:universality} provides our own derivation connecting the log derivatives of distance observables to the $\beta$-vector coefficients. 

The curvature analysis in Section~\ref{sec:curvature} connects to \citet{ChenZaldarriaga2025}, who find $\Omk = +0.0023 \pm 0.0011$ from standard MCMC methods. We did not run a dedicated $\Omega_k$ MCMC chain in this work; we use simple formulas to estimate the value of $\Omega_k$ that the BAO data prefer. 

Finally, \citet{Sailer2026} argue that a systematic in Planck's low-$\ell$ polarization $\tau$ would act as a single-point failure for essentially all current CMB analyses, and that the rest of the CMB$+$BAO data within \LCDM prefer $\tau = 0.090 \pm 0.012$, a $\sim 3$--$5\sigma$ shift from the low-$\ell$ measurement. Their $\tau$ channel and our $A_\mathrm{lens}$ channel involve the same physics expressed differently: $A_s e^{-2\tau}$ controls the CMB lensing amplitude, so $A_\mathrm{lens}$ and $\tau$ are partially degenerate. 

The paper is organized as follows. Section~\ref{sec:data_method} describes the data and SVD methodology. Section~\ref{sec:universal_c0} establishes the universality of \czero across all probes. Section~\ref{sec:c0_tensions} presents the \czero tension measurements. Section~\ref{sec:w0wa_reinterpretation} reinterprets the \wowa preference. Section~\ref{sec:curvature} examines spatial curvature. Section~\ref{sec:extensions} considers other CMB-side extensions. Section~\ref{sec:conclusions} discusses the implications and summarizes our conclusions. Appendix~\ref{app:universality} provides an analytical derivation of the \czero universality.

This paper, both calculations and draft, was written using Claude Code. It is based on some old calculations done at the time that the original DESI paper appeared and partially presented in a talk at Perimeter Institute\footnote{https://pirsa.org/25080017}. The goal of those calculations was just for me to better understand the claims and are presented here in case anyone finds them interesting. Creating this manuscript served as a guide for me to understand how to use the new AI tools. I make no claims of originality or novelty.  All code, data-loading scripts, and the full analysis pipeline needed to reproduce every number and figure in this paper are available in a public GitHub repository.\footnote{\url{https://github.com/matiaszaldarriaga/desi-w0wa-svd}} It contains the SVD analysis package, the scripts that turn the public DESI~DR2 BAO, supernova, and CMB MCMC chains into the distance-ratio decompositions used here, the marimo notebook that regenerates every figure, and a machine-readable registry of all claims and quoted numbers; the aim is that a reader can re-derive the results from the public inputs, inspect each numerical choice, and freely modify or extend the analysis. Appendix~\ref{app:reproducibility} describes the repository in more detail, and the reproducibility, number-provenance, and machine-readable-claim ideas it is meant to explore --- which may be of more interest than the specific results here. This manuscript will not be submitted to any journal. It has been sufficiently ``refereed" by Claude Code\footnote{This is a joke, but I am definitely of the camp that wants to let LLMs ``cook" (https://arxiv.org/abs/2602.10181)}. I am happy to correct any mistakes and of course anyone is free to download the code and ask any agent to fix or improve it. Hopefully this disclaimer is sufficient to appease the arxiv police\footnote{https://xcancel.com/tdietterich/status/2055055541542760694}, but maybe transparency will trigger a ban and you will have to wait a year for any new paper with my name on it. %

\section{Data and Method}
\label{sec:data_method}

\subsection{Data}
\label{sec:data}

We use three classes of low-redshift distance data, summarized in Table~\ref{tab:datasets}: DESI DR2 BAO measurements, three SN compilations (Union3, Pantheon+, DES-Dovekie), and the ACT DR6 / Planck 2018 CMB chains we use as the high-redshift reference. Each is described in turn below.

\subsubsection{DESI DR2 BAO}

We use the official DESI DR2 BAO measurements from \citet{DESI_DR2_BAO}. The data consist of 13 observables spanning $z = 0.295$ to $z = 2.330$: one isotropic measurement ($\DV/\rs$) from the Bright Galaxy Survey (BGS) at $z = 0.295$, plus six anisotropic redshift bins each contributing two observables ($\DM/\rs$ and $\DHubble/\rs$), for a total of $1 + 6 \times 2 = 13$ observables from Luminous Red Galaxies (LRG), Emission Line Galaxies (ELG), Quasars (QSO), and Lyman-$\alpha$ tracers.

The BAO observables are ratios of distances to the sound horizon. The comoving transverse distance is $\DM(z) = c\int_0^z dz'/H(z')$ (for the spatially flat models considered here; \Omk{} is reinstated in Section~\ref{sec:curvature}), the Hubble distance is $\DHubble(z) = c/H(z)$, and the volume-averaged distance is $\DV(z) = [z\,\DM(z)^2\,\DHubble(z)]^{1/3}$. For the supernovae the observable is the distance modulus $\mu(z) = 5\log_{10}[d_L(z)/10\,\mathrm{pc}]$, with luminosity distance $d_L(z) = (1+z)\,\DM(z)$. Here $\rs$ is the comoving size of the sound horizon at the baryon \emph{drag} epoch ($z_d \approx 1060$), the scale imprinted on the baryon distribution that calibrates the BAO feature \citep{DESI_DR2_BAO,ACT_DR6_extended}. It differs from the sound horizon at last scattering, $r_\star$ (sometimes written $r_{s,\star}$), which sets the CMB acoustic scale $\thetastar$; the two have different numerical values, and throughout we use the drag-epoch quantity $\rs$ to normalize the BAO distances. Our fiducial value is $\rs = 147.09$~Mpc.

The full $13 \times 13$ covariance matrix is block-diagonal in redshift --- measurements in different redshift bins are uncorrelated, while within each anisotropic bin $\DM/\rs$ and $\DHubble/\rs$ are jointly fit to the same 2D BAO feature and so are strongly anti-correlated ($r \approx -0.4$).

It is worth quantifying how tightly the CMB pins these distance ratios at high redshift. The acoustic scale $\thetastar = r_\star/\DM(z_\star)$, evaluated at the last-scattering redshift $z_\star$, is constrained to $0.025\%$ across the ACT \LCDM chain used below, so $\DM(z_\star)/r_\star$ is essentially fixed. The BAO observables differ only in normalizing by the drag-epoch horizon $\rs$ rather than $r_\star$ and---for $\DV$---in carrying an extra factor of the Hubble rate. Propagating these through the chain, $\DM(z_\star)/\rs$ has only $0.032\%$ scatter (the drag and last-scattering horizons are tightly correlated) and $\DV(z_\star)/\rs$ only $0.061\%$, both far below DESI's percent-level measurement errors. This is the quantitative reason the distance-ratio curves must return to the fiducial cosmology at high redshift, whatever the low-redshift expansion history.

\begin{deluxetable*}{lcccc}
\tablecaption{Dataset Summary \label{tab:datasets}}
\tablehead{
\colhead{Probe} & \colhead{$N_\mathrm{obs}$} & \colhead{$z$ range} & \colhead{Observable} & \colhead{Source}
}
\startdata
DESI DR2 BAO\tablenotemark{b} & 13 & $0.30$--$2.33$ & $\DV/\rs$, $\DM/\rs$, $\DHubble/\rs$ & \citet{DESI_DR2_BAO} \\
Union3 & 22 & $0.05$--$2.26$ & $\mu(z)$ & \citet{Union3} \\
Pantheon+ & 22\tablenotemark{a} & $0.04$--$2.12$ & $\mu(z)$ & \citet{PantheonPlus} \\
DES-Dovekie & 19\tablenotemark{a} & $0.04$--$1.14$ & $\mu(z)$ & \citet{Popovic2025} \\
\enddata
\tablenotetext{a}{Binned to Union3 22-bin grid; Pantheon+ and DES-Dovekie have 3 empty bins at $z > 1.12$.}
\tablenotetext{b}{See Table~\ref{tab:bao_observables} for the per-tracer breakdown.}
\end{deluxetable*}

\subsubsection{Type Ia Supernovae}

We analyze three SN compilations: Union3 \citep{Union3} with 22 pre-binned redshift bins ($z = 0.05$--$2.26$), Pantheon+ \citep{PantheonPlus} with 1701 individual SNe ($z = 0.001$--$2.26$), and DES-Dovekie \citep{Popovic2025} with 1755 SNe ($z = 0.025$--$1.14$). All cosmologically relevant information lies in a space that is smooth as a function of redshift, so we bin all datasets to Union3's common 22-bin grid for direct comparability---$\sim 20$ bins capture the full cosmological content. We bin using inverse-variance weighted averages. Let $i$ index the raw (unbinned) SN points and $j$ index the coarsened bins. With inverse-variance weights $w_i$ and bin weights $W_j = \sum_{i \in j} w_i$, the binning matrix $B_{ji} = w_i / W_j$ propagates the raw covariance to the binned one as $C_\mathrm{bin} = B\, C_\mathrm{raw}\, B^T$. To ensure binned data and binned model predictions are compared consistently, we evaluate chain predictions at effective redshifts $z_{\mathrm{eff},j}$ defined by $\mu_\mathrm{ref}(z_{\mathrm{eff},j}) = (B\,\mu_\mathrm{ref}(z_\mathrm{raw}))_j$, absorbing the bias from the nonlinearity of $\mu(z)$ across wide bins. After binning, all covariance matrices are well-behaved (positive-definite). Union3 has 22 occupied bins, Pantheon+ has 22, and DES-Dovekie has 19 (3 empty bins at $z > 1.14$). Calibration differences between datasets are absorbed by the absolute magnitude \MB marginalization (Section~\ref{sec:mb_marg}).

\subsubsection{CMB Chains}

We use ACT DR6 primary CMB (TT/TE/EE) + Planck 2018 low-$\ell$ (TT+EE) \LCDM chains \citep{ACT_DR6}, with no lensing, as our primary CMB reference, with 755,302 posterior samples from which we draw 2000 random samples for the SVD analysis.\footnote{Increasing to 5000 samples changes the normalized singular values by $< 4\%$ for the leading two modes, leaves all five $V$-vector inner products above 0.9999, and shifts the BAO \czero tension by $< 0.1\sigma$.} For beyond-\LCDM extensions, we use ACT chains with additional parameters (\Omk, $A_\mathrm{lens}$, $B_\mathrm{PMF}$, EDE $n=2$) \citep{ACT_DR6_extended}. The Planck 2018 chain \citep{Planck2018_cosmo} with 24,497 samples provides a cross-check (Table~\ref{tab:c0_tensions}).

We also use the official Planck + DESI DR2 \wowa{}CDM chain from \citet{DESI_DR2_cosmo}: Planck 2018 low-$\ell$ TT+EE + NPIPE high-$\ell$ CamSpec TTTEEE + DESI DR2 BAO, with no lensing, with ${\sim}91{,}000$ posterior samples from which we draw 2000. This chain carries the ${\sim}3\sigma$ dark energy preference that we discuss in Section~\ref{sec:w0wa_reinterpretation}.

\subsubsection{Reference Cosmology}
\label{sec:ref_cosmo}

All distance ratios and residuals are computed relative to the Planck 2018 \LCDM best-fit: $H_0 = 67.36$~km/s/Mpc, $\obh = 0.02237$, $\och = 0.1200$ ($\omh = 0.1424$), $\rs = 147.09$~Mpc.

\subsection{Motivation: Distance Ratio Plots}
\label{sec:motivation_plots}

\begin{figure}[t]
\centering
\includegraphics[width=\columnwidth]{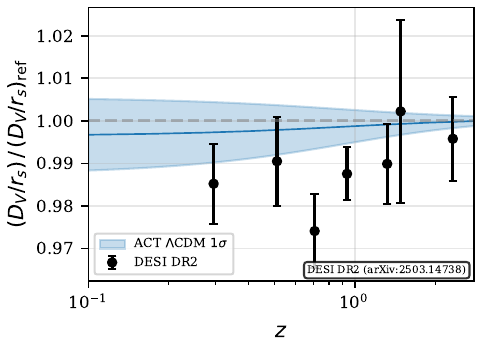}
\caption{DESI DR2 BAO volume-averaged distance ($\DV/\rs$) compared to ACT \LCDM predictions, shown as ratios relative to the Planck 2018 \LCDM reference. The blue band shows the $1\sigma$ range from the ACT chain. DESI data show systematic deviation at intermediate redshifts ($z \sim 0.5$--$1.5$). Note: $\DV$ is derived from $\DM$ and $\DHubble$ at each redshift for visualization only; the SVD analysis uses all 13 observables ($\DM/\rs$, $\DHubble/\rs$, $\DV/\rs$) separately (Table~\ref{tab:bao_observables}).}
\label{fig:dv_ratios}
\end{figure}

\begin{figure}[t]
\centering
\includegraphics[width=\columnwidth]{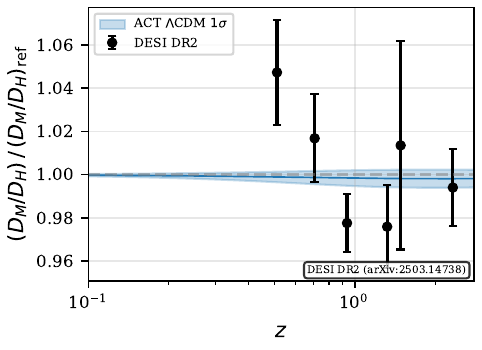}
\caption{Anisotropic distance ratio ($\DM/\DHubble$) for DESI DR2 BAO compared to ACT \LCDM predictions, relative to the Planck reference. The blue band shows the $1\sigma$ ACT range. This ratio is insensitive to the overall distance scale (both $\DM$ and $\DHubble$ share the same $\rs$ calibration). The lowest-redshift point deviates from the low-$z$ limit, the feature that would force a rapid evolution of $w(z)$ in a fit.}
\label{fig:dmdh_ratios}
\end{figure}

\begin{figure}[t]
\centering
\includegraphics[width=\columnwidth]{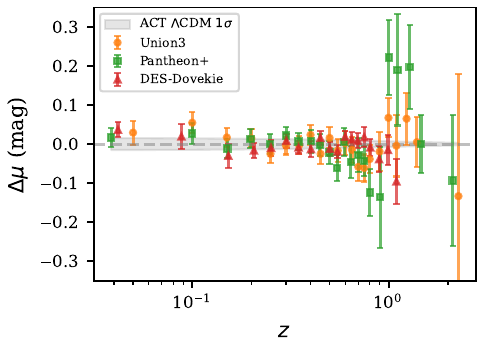}
\caption{All three SN datasets (mean-subtracted, binned to Union3 grid) vs.\ the ACT \LCDM $1\sigma$ prediction envelope. Red circles: Union3 (22 bins). Blue squares: Pantheon+ (22 bins). Green triangles: DES-Dovekie (19 bins). The ACT band is narrow ($\Delta\mu \lesssim 0.01$~mag) because ACT tightly constrains \omh.}
\label{fig:sn_datasets}
\end{figure}

Before developing the SVD framework, we preview the raw data-vs-model comparison to motivate the analysis (Figures~\ref{fig:dv_ratios}--\ref{fig:sn_datasets}).

Key observations: DESI $\DV/\rs$ measurements lie systematically below the CMB \LCDM prediction at $z \sim 0.5$--$1.5$. All ratio curves converge to unity at high $z$ because \thetastar is so well measured by the CMB. The SVD analysis (Section~\ref{sec:svd_method}) will formalize what these plots show qualitatively.

These data overviews (Figures~\ref{fig:dv_ratios}--\ref{fig:sn_datasets}) already reveal the key features that the SVD analysis will quantify. First, DESI $\DV/\rs$ measurements lie systematically below the CMB \LCDM prediction at $z \sim 0.5$--$1.0$: the Hubble rate must be higher than the reference cosmology at low redshifts to get the shorter distances. But $\thetastar$ fixes the distance to the last-scattering surface, so $H(z)$ must eventually drop below the reference at higher redshift to compensate---a requirement that, within the dark energy framework, implies $w < -1$ at some epoch. This is so because within the dark energy framework, the dark energy density must increase with time relative to the cosmological constant in LCDM, forcing the model to cross the phantom divide. Of course if we think of the combined dark-matter dark-energy fluid, whose equation of state is not close to $-1$, no such requirement exists.

Second, we also plot the ratio $(\DM/\DHubble)/(\DM/\DHubble)_\mathrm{ref}$ (Figure~\ref{fig:dmdh_ratios}), which approaches unity at low $z$ for any cosmology where the Hubble parameter can be considered constant in that limit. In that case, both terms in the ratio reduce to $z$, so deviations from unity at low $z$ are small. The lowest DESI data point at $\DM/\DHubble = 1.04$ deviates from this limit, forcing a rapid evolution of $w(z)$ if one wants to fit this point. 

Third, the SN plots show that they have substantially less constraining power than BAO: the error bars on the SN residuals are visibly wider than the $1\sigma$ ACT $\Lambda$CDM band.

\subsection{SVD Methodology}
\label{sec:svd_method}

Throughout this section, $s = 1,\ldots,N$ indexes chain samples (we use $N = 2000$), $i = 1,\ldots,D$ indexes data observables (so $D = 13$ for BAO, $22$ for Union3 and Pantheon+, $19$ for DES-Dovekie), and $\alpha = 0, 1, \ldots, D-1$ indexes SVD modes ordered by decreasing singular value.

We adopt one convention throughout. We call the $V_\alpha$ (the columns of $V$) \emph{modes} or \emph{directions}: unit vectors in whitened observable space. The scalar projection of a chain sample onto mode $\alpha$ is its \emph{chain amplitude} $c_\alpha^{(s)}$, and the projection of the data onto the same direction is the \emph{data amplitude} $a_\alpha$. When we write \czero, $c_1$, \ldots\ without a sample superscript we mean these amplitudes (the data amplitude when comparing to data, the chain mean when comparing to the model); the directions themselves are always written $V_0, V_1, \ldots$

\subsubsection{BAO: Ratio Space}

For each CMB chain sample, we compute the 13 BAO observables listed in Table~\ref{tab:bao_observables} at DESI redshifts, using that sample's own cosmological parameters---including its sample-specific sound horizon \rs. We form ratios of each chain prediction relative to the reference cosmology: $\delta_i^{(s)} = \mathrm{pred}_i^{(s)} / \mathrm{ref}_i - 1$, where $\mathrm{pred}_i^{(s)}$ is the predicted observable for chain sample $s$ and $\mathrm{ref}_i$ uses the Planck best-fit parameters (Section~\ref{sec:ref_cosmo}). The DESI observational covariance is transformed to ratio space: $C_\mathrm{ratio}(i,j) = C_\mathrm{abs}(i,j) / (\mathrm{ref}_i \times \mathrm{ref}_j)$.

\begin{deluxetable}{cllc}
\tablecaption{The 13 DESI DR2 BAO observables used in the SVD analysis. Each anisotropic redshift bin contributes two observables ($\DM/\rs$ and $\DHubble/\rs$); the lowest bin provides only $\DV/\rs$.\label{tab:bao_observables}}
\tablehead{
\colhead{$z_\mathrm{eff}$} & \colhead{Tracer} & \colhead{Observables} & \colhead{$N_\mathrm{obs}$}
}
\startdata
0.295 & BGS        & $\DV/\rs$                    & 1 \\
0.510 & LRG1       & $\DM/\rs$, $\DHubble/\rs$    & 2 \\
0.706 & LRG2       & $\DM/\rs$, $\DHubble/\rs$    & 2 \\
0.934 & LRG3+ELG1  & $\DM/\rs$, $\DHubble/\rs$    & 2 \\
1.321 & ELG2       & $\DM/\rs$, $\DHubble/\rs$    & 2 \\
1.484 & QSO        & $\DM/\rs$, $\DHubble/\rs$    & 2 \\
2.330 & Ly$\alpha$ & $\DM/\rs$, $\DHubble/\rs$    & 2 \\
\enddata
\end{deluxetable}

\subsubsection{SN: Difference Space with \MB Marginalization}
\label{sec:mb_marg}

For each chain sample, we compute distance modulus residuals: $\Delta\mu_i = \mu_\mathrm{model}(z_i) - \mu_\mathrm{ref}(z_i)$. The unknown absolute magnitude \MB is marginalized by inflating the covariance:
\begin{equation}
C_\mathrm{marg} = C_\mathrm{bin} + \sigma_M^2 \cdot \mathbf{1}\mathbf{1}^T
\label{eq:mb_marg}
\end{equation}
with $\sigma_M = 100$ (in magnitudes, vastly larger than any plausible $M_B$ uncertainty; in the $\sigma_M \to \infty$ limit this is mathematically equivalent to projecting out the constant-offset direction, but a finite value keeps the covariance well-conditioned), adding one very large eigenvalue ($\approx N \sigma_M^2$) in the constant-offset direction. After whitening, this direction gets weight $\sim 1/\sqrt{N\sigma_M^2} \approx 0.002$ for $N = 22$---effectively zero. The result is stable for $\sigma_M \geq 5$.

\subsubsection{Whitening}

We whiten using the eigendecomposition of the (ratio-transformed or \MB-marginalized) covariance: $\tilde{y} = C^{-1/2} y$. Whitening puts all directions on equal footing: in $\tilde y$-space every linearly independent combination of the observables has unit measurement variance, so the subsequent SVD picks out directions of largest \emph{model} variation across the CMB posterior, not directions that happen to have large measurement errors. In the whitened space, every direction has unit measurement variance. For SN datasets, the \MB marginalization (Eq.~\ref{eq:mb_marg}) inflates one eigenvalue to $\sim N\sigma_M^2 \approx 2.2 \times 10^5$; whitening assigns that direction negligible weight, effectively marginalizing \MB without discarding the dimension. All covariances are well-conditioned after binning, so all dimensions are retained: 13 for BAO, 22 for Union3 and Pantheon+, and 19 for DES-Dovekie (whose recalibrated covariance \citep{Popovic2025} is well-conditioned).

\subsubsection{SVD}

The $N \times D$ matrix of whitened chain predictions ($N = 2000$ samples, $D$ the data dimension) is decomposed as $M = U S V^T$. We do \emph{not} mean-center $M$ before the SVD. The offset of the chain centroid from the reference cosmology is itself a physical signal --- it carries the difference between the reference cosmology's predictions and what the CMB chain prefers, and discarding it could discard the very tension we are trying to measure. For BAO, $\langle \tilde y \rangle$ is therefore generically nonzero; for SN, the $M_B$ marginalization already absorbs any constant-offset direction (Section~\ref{sec:mb_marg}), so the distinction is immaterial. The columns of $V$ are the principal directions ordered by singular value $S_\alpha$.

Chain coefficients: $c_\alpha^{(s)} = U_{s\alpha} S_\alpha$ gives the projection of chain sample $s$ onto mode $\alpha$. The chain mean $\langle c_\alpha \rangle$ and standard deviation $\sigma_{c_\alpha}$ over the $N$ samples characterize the CMB posterior's location and spread in direction $\alpha$.
Data projection: $a_\alpha = (\tilde{y}_\mathrm{data}) \cdot V_\alpha$. Because the $V_\alpha$ are orthonormal and the data are whitened, the measurement error on $a_\alpha$ is 1 in whitened units for every $\alpha$ --- every mode carries the same measurement noise by construction.

The relevant quantity is the \emph{standard deviation}:
\begin{equation}
\sigma_{c_\alpha} \equiv \mathrm{std}(c_\alpha)
\label{eq:sigma_c}
\end{equation}
which measures how much the model prediction moves across the CMB posterior in direction $\alpha$, in units of measurement error. This is what we report throughout the paper. The interpretation is simple:
\begin{itemize}
\item $\sigma_{c_\alpha} > 1$: the low-$z$ probe constrains direction $\alpha$ more tightly than the CMB. New information flows from the probe to the CMB.
\item $\sigma_{c_\alpha} < 1$: the CMB already constrains direction $\alpha$ better than the probe can measure it.
\end{itemize}

The key point is that the SVD modes and their singular values are determined entirely by the CMB chains and the low-$z$ covariance matrix---\emph{before looking at the low-$z$ data}. The data only enter when we project onto these predetermined directions.

\subsubsection{Tension Formula}

The tension between data and model in mode $\alpha$ is defined below. The numerator measures how far the data sits from the chain mean in mode $\alpha$; the denominator combines, in quadrature, the unit measurement variance (from whitening) and the chain variance $\sigma_{c_\alpha}^2$ (the model uncertainty across the CMB posterior). Both variances are in the same whitened units.
\begin{equation}
\mathrm{tension}_\alpha = \frac{a_\alpha - \langle c_\alpha \rangle}{\sqrt{1 + \sigma_{c_\alpha}^2}}
\label{eq:tension}
\end{equation}
where $a_\alpha$ is the data projection, $\langle c_\alpha \rangle$ is the chain mean, and $\sigma_{c_\alpha}$ is the chain standard deviation (Eq.~\ref{eq:sigma_c}). With this convention, positive tension means the data amplitude $a_\alpha$ lies above the chain mean $\langle c_\alpha \rangle$. The ``1'' in the denominator is the unit measurement variance in whitened space (by construction). Thus the denominator combines two independent uncertainties: observational measurement error (always 1 in whitened units) and model uncertainty from the CMB posterior ($\sigma_{c_\alpha}^2$). This formula assumes Gaussian distributions for the $c_\alpha$ coefficients; Kolmogorov--Smirnov tests confirm that the leading-mode distributions are consistent with Gaussian for all four probes.

\subsubsection{Parameter Formulas via $\beta$-vectors}

To identify the physical meaning of each SVD mode, we regress the sigma-normalized SVD coefficient against sigma-normalized cosmological parameters:
\begin{equation}
\frac{c_\alpha - \langle c_\alpha \rangle}{\sigma_{c_\alpha}} = \sum_p \beta_p^\alpha \frac{X_p - \langle X_p \rangle}{\sigma_{X_p}}
\label{eq:beta_vectors}
\end{equation}
where $X_p \in \{\omh, \obh, \thetastar\}$. Because both sides are standardized, the $\beta$ coefficients reflect the product of physical sensitivity and CMB posterior width: a parameter that the CMB constrains loosely will have a larger $|\beta|$ for a given physical sensitivity. The $\beta$-vector coefficients reveal which combination of CMB parameters drive each mode.

\subsubsection{Extension SVD}
\label{sec:extension_svd_method}

When additional parameters are varied (e.g., \wowa, \Omk, $A_\mathrm{lens}$), the model predictions sweep a larger volume of observable space. The key question is whether this additional freedom opens \emph{new measurable directions} beyond \czero, or whether it merely expands the range of motion along the same direction.

To search for genuinely new directions, that are a priori reasonable given the model, we project out \czero from the whitened chain predictions and decompose the residuals:
\begin{equation}
y_\perp = y - (y \cdot V_0)\, V_0
\end{equation}
We then perform SVD on $y_\perp$ to obtain residual modes. The correct measurability criterion is $\sigres \equiv \mathrm{std}(c_\mathrm{res})$, the standard deviation of the leading residual coefficient across the chain posterior. This measures how much the model prediction moves in the new direction, in units of measurement error. As a rule of thumb, we will require $\sigres > 0.3$ for a direction to be considered measurable: below this threshold, the model displacement is less than $1/3$ of a measurement $\sigma$, contributing $< 0.1$ to $\chi^2$ in the new direction and making the mode indistinguishable from noise. Of course this is an arbitrary threshold only used to guide our intuition.
\dataref{method_svd}{calculation/scripts/svd\_lcdm\_analysis.py}

\section{The Universal Mode}
\label{sec:universal_c0}

\claim{c0_universal}{All four distance probes define the same leading SVD direction $V_0$, whose data amplitude is \czero (cross-probe inner products $> 0.996$).}
\claim{c0_is_omegamh2}{\czero is a near-perfect proxy for \omh ($R^2 > 0.99$).}

The SVD of whitened chain predictions reveals that one mode dominates for every probe. Table~\ref{tab:sigma_c} summarizes the per-mode standard deviations $\sigma_{c_\alpha}$ (Eq.~\ref{eq:sigma_c}).

\begin{deluxetable}{lcccc}
\tablecaption{Leading Two SVD Mode Standard Deviations $\sigma_{c_\alpha}$ by Probe \label{tab:sigma_c}}
\tablehead{
\colhead{Mode} & \colhead{BAO} & \colhead{Union3} & \colhead{Pantheon+} & \colhead{DES-Dovekie}
}
\startdata
$\sigma_{c_0}$ & 1.56 & 0.28 & 0.30 & 0.33 \\
$\sigma_{c_1}$ & 0.063 & 0.0010 & 0.0011 & 0.0009 \\
$\sigma_{c_0}/\sigma_{c_1}$ & 25 & 273 & 275 & 355 \\
\enddata
\end{deluxetable}

Within \LCDM, only the amplitude of the leading mode is measurable for any of the four probes. BAO has $\sigma_{c_0} = 1.56 > 1$: the CMB chain populates the $V_0$ direction by more than one BAO measurement error, so projecting the BAO data onto $V_0$ constrains \omh more tightly than the CMB itself. The three SN datasets have $\sigma_{c_0} \approx 0.28$--$0.33 < 1$: the CMB already constrains $V_0$ tighter than the SN data can --- projecting the SN data onto $V_0$ adds little information about \omh beyond the CMB prior. No probe has a second measurable \LCDM direction ($\sigma_{c_1} \ll 1$ for all, including BAO at $\sigma_{c_1} = 0.06$).

\subsection{Universality and Physical Interpretation}
\label{sec:c0_interpretation}

All four probes define almost the same \czero direction in parameter space. Table~\ref{tab:beta_vectors} shows the $\beta$-vector coefficients (Eq.~\ref{eq:beta_vectors}) for each probe:

\begin{deluxetable}{lccc}
\tablecaption{$\beta$-Vector Coefficients for \czero. In all cases the three-parameter fit achieves $R^2 > 0.999$. \label{tab:beta_vectors}}
\tablehead{
\colhead{Probe} & \colhead{$\beta_{\omh}$} & \colhead{$\beta_{\obh}$} & \colhead{$\beta_{\thetastar}$}
}
\startdata
BAO (DESI) & $-0.90$ & $+0.13$ & $+0.17$ \\
Union3 & $-0.90$ & $+0.18$ & $+0.12$ \\
Pantheon+ & $-0.90$ & $+0.18$ & $+0.12$ \\
DES-Dovekie & $-0.90$ & $+0.18$ & $+0.12$ \\
\enddata
\end{deluxetable}

\evidence{c0_universal}{ev_inner_products}{Inner product matrix: minimum = 0.9969, mean = 0.9984 across all probe pairs.}

To compare what the four probes' leading directions $V_0$ have in common, we must first express them in a common space. Each probe's $V_0$ is a unit vector of length $D_\mathrm{probe}$ in its own whitened observable space (e.g., 13-D for BAO, 22-D for Union3). We map each $V_0$ into a common 3-D CMB parameter space via the $\beta$-vector regression (Eq.~\ref{eq:beta_vectors}): the unit-norm vector $\hat\beta = (\beta_{\Omega_m h^2}, \beta_{\Omega_b h^2}, \beta_{\theta_\star})/\|\beta\|$ identifies the linear combination of CMB parameters that each probe's $V_0$ is sensitive to. The pairwise dot product $\hat\beta^{(A)} \cdot \hat\beta^{(B)}$ is what we call the cross-probe inner product. A value of 1 means the two probes' $V_0$ directions correspond to identical parameter combinations; 0 would mean orthogonal. The cross-probe inner product for \czero exceeds 0.996 for every pair. The three SN datasets define virtually identical \czero directions; BAO differs slightly due to its sensitivity to the sound horizon \rs, but the BAO--SN inner product still exceeds 0.996. This universality is not guaranteed a priori: BAO measures $D/\rs$ ratios while SN measures distance moduli $\mu(z)$ --- at low redshift these depend on different combinations of cosmological parameters. The convergence to the same direction is a consequence of the CMB \LCDM posterior: once the CMB pins down the parameters the residual freedom in low-redshift distance predictions runs almost entirely along one direction, which we loosely call the $\Omega_m h^2$ direction, regardless of which probe is doing the measuring.

The dominance of \omh indicates that \czero is primarily an \omh proxy. Here $R^2$ is the fraction of the chain-sample variance of $c_0$ explained by a linear regression on the listed CMB parameters; $R^2 = 1$ would mean $c_0$ is exactly a linear combination of those parameters across the CMB posterior. Sequential multi-parameter fits quantify this:
\begin{itemize}
\item \omh alone: $R^2 = 0.95$
\item $\omh + \obh$: $R^2 = 0.974$
\item $\omh + \obh + \thetastar$: $R^2 = 0.99998$
\end{itemize}
The first line is from a practical perspective the non-trivial content: \omh alone explains 95\% of the chain variance in \czero. Adding \obh and \thetastar absorbs the residual 5\%, giving $R^2 = 0.99998$. The three-parameter $R^2 \approx 1$ is expected --- the remaining \LCDM parameters (optical depth, primordial amplitude and tilt) do not enter low-redshift distances at the linear level --- but the 0.95 from \omh alone is what makes \czero usable as an \omh proxy. For brevity we will refer to \czero as the \omh mode throughout, with the understanding that it is the specific linear combination of (\omh, \obh, \thetastar) along the CMB degeneracy direction.

\evidence{c0_is_omegamh2}{ev_beta_regression}{Sequential fit: \omh alone $R^2=0.95$, $+\obh$ $R^2=0.974$, $+\thetastar$ $R^2=0.99998$. Beta regression on all four probes.}

The sigma-normalized formula for BAO is:
\begin{multline}
\frac{\czero - \langle \czero \rangle}{\sigma_{\czero}} = -0.90\, \frac{\omh - \langle \omh \rangle}{\sigma_{\omh}} \\
+ 0.13\, \frac{\obh - \langle \obh \rangle}{\sigma_{\obh}} + 0.17\, \frac{\thetastar - \langle \thetastar \rangle}{\sigma_{\thetastar}}
\label{eq:c0_formula}
\end{multline}
The SN $\beta$-vectors are nearly identical (Table~\ref{tab:beta_vectors}), so the sigma-normalized \czero is the same linear combination of cosmological parameters for every probe. 

The difference between probes lies only in the normalization: $\czero = 1$ corresponds to $1\sigma$ of measurement error for that probe, and since BAO and SN have very different measurement precisions, a unit of BAO \czero and a unit of SN \czero correspond to different shifts in \omh. Specifically, because the $\beta$-vectors are almost proportional, the \czero values for any two probes are related by:
\begin{equation}
c_0^\mathrm{SN} \approx \frac{\sigma_{c_0}^\mathrm{SN}}{\sigma_{c_0}^\mathrm{BAO}} \, c_0^\mathrm{BAO}
\label{eq:c0_sn_bao}
\end{equation}
For DES-Dovekie, $\sigma_{c_0}^\mathrm{SN}/\sigma_{c_0}^\mathrm{BAO} = 0.33/1.56 \approx 0.21$: a given cosmological shift produces a $\sim 5\times$ smaller \czero value in SN whitened units than in BAO whitened units, reflecting the lower sensitivity of SN to \omh.

The dominance of \omh follows quantitatively from the logarithmic derivatives of low-redshift distance observables combined with the CMB parameter covariance (Appendix~\ref{app:universality}). We define the \emph{per-sigma sensitivity} of an observable to a parameter $p$ as the fractional change in the observable per $1\sigma$ shift in $p$ along the CMB posterior --- i.e., the log derivative times the CMB fractional error. For $\DM(z)/r_d$ and \omh, this is $\sim 5\times$ larger than for \obh, the product of (i) a log derivative $\partial \ln(\DM/r_d)/\partial \ln \omh \approx 0.8$ which is $\sim 3\times$ larger than $|\partial\ln(\DM/r_d)/\partial\ln \obh| \approx 0.3$, and (ii) a CMB fractional error on \omh of $0.8\%$ which is $\sim 1.6\times$ larger than the $0.5\%$ on \obh. The large \thetastar derivative ($g_{\thetastar} \approx -4$) is suppressed by the tiny fractional error ($0.025\%$). 

The correlations between parameters in the CMB posterior further amplify the \omh dominance: when regressing on \omh alone, the correlated parts of \obh and \thetastar are partially captured, so that \omh alone predicts $R^2 = 0.95$. Appendix~\ref{app:universality} predicts the $\beta$-coefficients in Table~\ref{tab:beta_vectors} to within 2\% for \omh and ${\sim}\,30\%$ for \obh from single-redshift proxy observables. The results in the appendix are presented simply to gain some intuition about the results. 

For SN, the $M_B$ marginalization removes the absolute distance scale, leaving the \emph{shape} of the distance--redshift relation. The relevant proxy is the distance ratio $\DM(0.3)/\DM(0.5)$, which has no $r_d$ dependence and no $M_B$ ambiguity. Its predicted $\beta$ is similar to the BAO proxy's (Appendix~\ref{app:universality}, Table~\ref{tab:beta_comparison}), explaining why the SN $\beta$-vectors in Table~\ref{tab:beta_vectors} nearly match the BAO values. The small difference ($\beta_{\obh} = +0.18$ for SN vs $+0.13$ for BAO) traces to the fact that BAO has an additional \obh sensitivity through $r_d$ that the SN distance ratio lacks.

The MEDI framework \citep{Weiner2026_MEDI} provides the analytic counterpart to this result: it identifies $\omega_m r_d^2$ --- rather than \omh alone --- as the parameter combination that acoustic-scale measurements (BAO calibrated by the CMB) are fundamentally sensitive to. Within \LCDM the drag-epoch sound horizon $r_d$ is nearly fixed by early-universe physics, so $\omega_m r_d^2$ is to good approximation proportional to \omh; this is why \czero emerges as essentially an \omh mode, and also why regressing \czero on $\omega_m r_d^2$ across the \LCDM chain is no more informative than regressing on \omh alone --- the chain does not exercise $r_d$ enough to separate the two. The value of the MEDI identification is that it holds analytically, including in extended models where $r_d$ can shift. See the introduction for a fuller discussion of the relation to previous work.

\begin{figure*}[t]
\centering
\includegraphics[width=\textwidth]{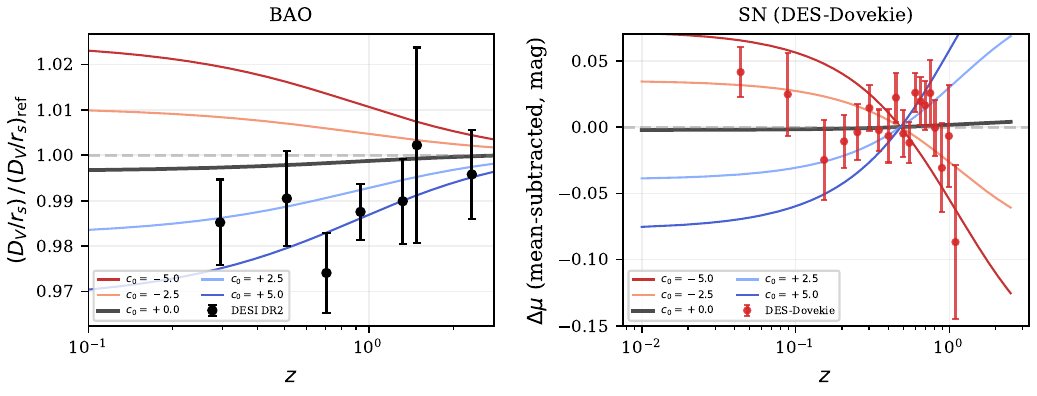}
\caption{\czero family of curves in observed space. \emph{Left:} BAO $D_V/\rs$ ratio (normalized to the Planck reference cosmology) for five values of \czero spanning $\pm 5$ whitened units ($\approx \pm 3\sigma$ of the ACT chain). Black points: DESI DR2. \emph{Right:} SN distance modulus residuals $\Delta\mu$ (mean-subtracted) for the same \czero range. Points: DES-Dovekie. The \czero normalization differs between panels (Eq.~\ref{eq:c0_sn_bao}): BAO \czero $= 5$ corresponds to a $\sim 5\times$ larger \omh shift than SN \czero $= 5$.}
\label{fig:c0_families}
\end{figure*}

Figure~\ref{fig:c0_families} shows what varying \czero looks like in observable space. The left panel displays BAO $D_V/\rs$ ratios (normalized to the Planck reference cosmology) for five values of \czero spanning $\pm 5$ in BAO whitened units, which corresponds to $\pm 5 / \sigma_{c_0}^\mathrm{BAO} \approx \pm 3.2$ in units of the ACT \LCDM chain spread in $V_0$ (Table~\ref{tab:sigma_c}); the right panel shows the corresponding SN distance modulus residuals (mean-subtracted to marginalize $M_B$). The \czero values in the two panels do not correspond to the same cosmological parameters: the normalization of \czero is set by each probe's measurement covariance (Eq.~\ref{eq:c0_sn_bao}), so $c_0 = 5$ represents a $\sim 5\times$ larger shift in \omh for BAO than for SN.

\dataref{ev_inner_products}{calculation/scripts/cross\_dataset\_analysis.py}

\section{\czero Tensions: \omh from All Probes}
\label{sec:c0_tensions}

\claim{c0_tensions_sign}{BAO alone prefers lower \omh ($+2.2\sigma$). All three SN datasets are mild or null: Union3 $-1.7\sigma$, Pantheon+ $-1.1\sigma$, DES-Dovekie $-0.8\sigma$ (all higher \omh or consistent with zero).}
\depends{c0_tensions_sign}{c0_universal}
\claim{bao_constrains_omh2}{BAO constrains \omh more tightly than the CMB ($\sigma_{c_0} = 1.56 > 1$). SN datasets do not ($\sigma_{c_0} \approx 0.28$--$0.33 < 1$).}
\depends{bao_constrains_omh2}{c0_universal}

Since all probes measure the same \czero direction (Section~\ref{sec:universal_c0}), we can compare their \czero measurements directly. The tension formula (Eq.~\ref{eq:tension}) gives four independent measurements of the same quantity, although we need to remember that the \czero normalization differs between probes. A unit shift in BAO \czero and a unit shift in SN \czero both correspond to $1\sigma$ of their respective measurement errors, but these translate to different shifts in \omh (see Section~\ref{sec:money_plot}).

Table~\ref{tab:c0_tensions} summarizes the results. A few points of interpretation are needed before examining the numbers:

\begin{itemize}
\item \textbf{Sign convention:} Positive tension means the data projection $a_0$ exceeds the chain mean $\langle c_0 \rangle$. Since $\beta_{\omh} = -0.90$, a positive \czero shift corresponds to \emph{lower} \omh (larger distances).

\item \textbf{What $\sigma_{c_0}$ means:} This is the standard deviation of \czero across the CMB chain posterior---it measures how much the CMB model prediction varies in the \czero direction, in units of measurement error. It is \emph{not} the measurement error, which is always 1 in whitened space by construction. The tension denominator $\sqrt{1 + \sigma_{c_0}^2}$ combines both.

\item \textbf{$\sigma_{c_0}$ and constraining power:} BAO has $\sigma_{c_0} = 1.56 > 1$. This means BAO constrains \omh more tightly than the CMB: DESI data carries genuinely new information about \omh beyond what the CMB already provides. The three SN datasets have $\sigma_{c_0} \approx 0.28$--$0.33 < 1$: the CMB already provides a tighter constraint than these SN measurements, so their \czero tensions are measured against a prior that is already narrower than the data.
\end{itemize}

\begin{deluxetable*}{lcccc}
\tablecaption{\czero Tension per Probe \label{tab:c0_tensions}}
\tablecomments{Columns: $\langle c_0 \rangle$ is the mean of the CMB chain's \czero distribution (the model prediction in the $V_0$ direction); $\sigma_{c_0}$ its standard deviation across the chain (model uncertainty, in units of the measurement error); $a_0$ is the data amplitude, the projection of the measured low-$z$ data onto $V_0$ (measurement error 1 by construction); and the tension is $(a_0 - \langle c_0 \rangle)/\sqrt{1+\sigma_{c_0}^2}$ (Eq.~\ref{eq:tension}). Values in parentheses use the Planck 2018 \LCDM chain as the CMB reference instead of ACT DR6.}
\tablehead{
\colhead{Probe} & \colhead{$\langle c_0 \rangle$} & \colhead{$\sigma_{c_0}$} & \colhead{$a_0$} & \colhead{Tension}
}
\startdata
BAO (DESI) & $+0.59$ ($-0.31$) & 1.56 (1.85) & $+4.69$ & $+2.2\sigma$ ($+2.4\sigma$) \\
Union3 & $+0.13$ ($-0.06$) & 0.28 (0.34) & $-1.59$ & $-1.7\sigma$ ($-1.4\sigma$) \\
Pantheon+ & $+0.14$ ($-0.06$) & 0.30 (0.36) & $-0.97$ & $-1.1\sigma$ ($-0.9\sigma$) \\
DES-Dovekie & $+0.16$ ($-0.07$) & 0.33 (0.40) & $-0.74$ & $-0.8\sigma$ ($-0.6\sigma$) \\
\enddata
\end{deluxetable*}

\evidence{c0_tensions_sign}{ev_c0_table}{Four-probe tension table: BAO positive, all SN mild/negative.}

BAO prefers lower \omh than the CMB (positive \czero tension at $+2.2\sigma$), while the three SN datasets prefer higher \omh --- Union3 at $-1.7\sigma$, Pantheon+ at $-1.1\sigma$, and DES-Dovekie at $-0.8\sigma$. None of the SN tensions is individually significant, and because $\sigma_{c_0}^\mathrm{SN} < 1$ the SN datasets do not have the constraining power on \czero to confirm or refute the BAO result; the opposite-sign pattern is suggestive but should not be interpreted as a measured disagreement.

\begin{figure}[t]
\centering
\includegraphics[width=\columnwidth]{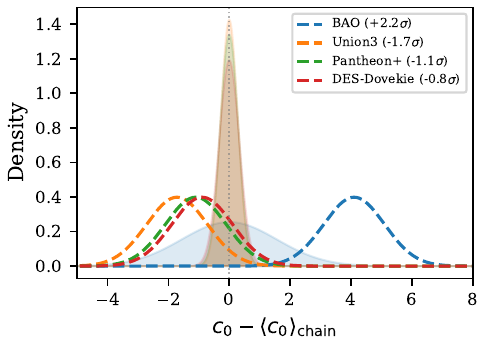}
\caption{Per-probe \czero distributions. \emph{Shaded regions}: CMB chain predictions (model uncertainty in each probe's \czero). \emph{Dashed curves}: measurement likelihoods centered at the data \czero value ($\sigma = 1$ in whitened space by construction). The BAO chain distribution (shaded, wide) has $\sigma_{c_0} = 1.56$, wider than the measurement likelihood, reflecting BAO's ability to constrain \omh beyond the CMB prior. The three SN chain distributions (shaded, narrow, $\sigma_{c_0} \approx 0.28$--$0.33$) are much narrower than their measurement likelihoods, indicating SN data adds little constraining power on \czero.}
\label{fig:c0_gaussians}
\end{figure}

An important point to remember when interpreting Figure~\ref{fig:c0_gaussians}: the \czero normalization differs between probes. A unit shift in BAO \czero and a unit shift in SN \czero both correspond to $1\sigma$ of their respective measurement errors, but these translate to different shifts in \omh (see Section~\ref{sec:money_plot}). The figure is designed to visualize \emph{tensions}, not to compare \omh values directly. 

\subsection{The \omh Money Plot}
\label{sec:money_plot}

Since \czero is a near-perfect proxy for \omh (Eq.~\ref{eq:c0_formula}), we can convert each probe's \czero measurement to an implied \omh value. This is the value of \omh implied by the data if one slides along the CMB posterior's degeneracy direction in $(\omh, \obh, \thetastar)$. The conversion inverts the $\beta$-vector formula, setting \obh and \thetastar to their chain mean values:
\begin{equation}
\omh_\mathrm{probe} = \langle \omh \rangle_\mathrm{chain} + (a_0 - \langle c_0 \rangle) \times \frac{\sigma_{\omh}}{\beta_{\omh} \cdot \sigma_{c_0}}
\label{eq:omh_conversion}
\end{equation}
where $\beta_{\omh}$ is the $\beta$-vector coefficient for \omh, and $\sigma_{\omh}$, $\sigma_{c_0}$ are the chain standard deviations. The measurement error on \omh from the $\sigma = 1$ whitened-space uncertainty is $\sigma_{\omh}^\mathrm{meas} = |\sigma_{\omh} / (\beta_{\omh} \cdot \sigma_{c_0})|$. An additional marginalization uncertainty arises from the CMB posterior widths of \obh and \thetastar; this adds $\sigma_\mathrm{marg} \approx 0.0003$ in quadrature, a modest correction ($\sim 7\%$ for BAO, negligible for SN).

\begin{figure}[t]
\centering
\includegraphics[width=\columnwidth]{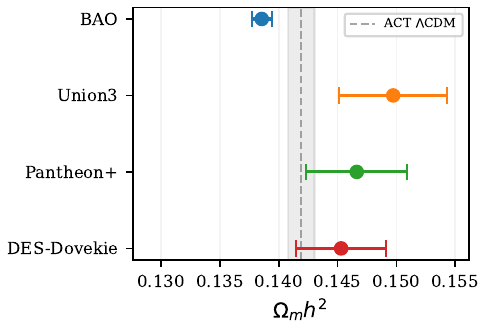}
\caption{Effective \omh measurements from all four distance probes, derived from \czero via the $\beta$-vector inversion (Eq.~\ref{eq:omh_conversion}). Error bars show total uncertainty (measurement $+$ marginalization over \obh and \thetastar, added in quadrature). The gray band shows the ACT \LCDM posterior width. BAO provides a tighter constraint than the CMB.}
\label{fig:omegamh2_money}
\end{figure}

Figure~\ref{fig:omegamh2_money} shows the result. BAO provides the tightest \omh constraint of the four probes --- tighter than the CMB itself --- while the three SN measurements have total errors $\sim 5\times$ larger.

\subsection{Physical Interpretation of the Mode Shape}

\begin{figure*}[t]
\centering
\includegraphics[width=\textwidth]{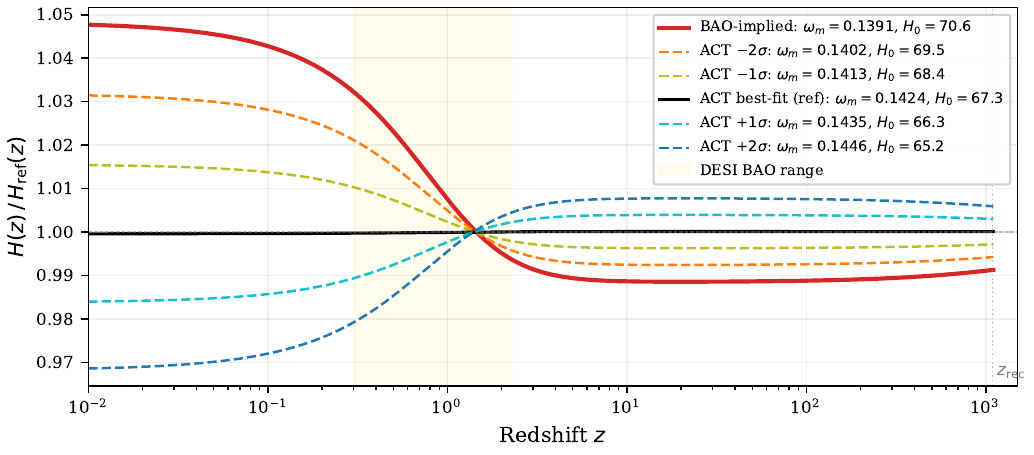}
\caption{Hubble rate ratio $H(z)/H_\mathrm{ref}(z)$ for varying \omh at fixed \thetastar (flat \LCDM). Higher \omh (blue, dashed) raises $H$ at high $z$ but lowers $H_0$; the crossover near $z \sim 2$ falls within the DESI BAO range (shaded). The red curve shows the BAO-implied \omh from the \czero measurement ($\omh = 0.139$, $H_0 = 70.6$).}
\label{fig:hz_omh2}
\end{figure*}

The \czero mode measures \omh, and the physical effect of varying \omh at fixed \thetastar is a characteristic tilt of the Hubble rate $H(z)$ (Figure~\ref{fig:hz_omh2}). Increasing \omh raises $H(z)$ at high redshift (where $H \propto \sqrt{\omh}$ in the matter era), but the \thetastar constraint forces $H_0$ to \emph{decrease} to compensate (keeping $r_\star / D_A(z_\mathrm{rec})$ fixed). The result is a crossover near $z \sim 1$--$2$: higher \omh means faster expansion at $z > 2$ but slower at $z < 1$. This crossover falls squarely in the DESI BAO redshift range, explaining why BAO is sensitive to \omh.

For distances, lower \omh (positive \czero) produces smaller $D_V/\rs$ ratios at all BAO redshifts (Figure~\ref{fig:c0_families}, left panel): the slower high-$z$ expansion reduces distances, while the higher $H_0$ reduces $\rs$-normalized quantities further. For SN, the same physics manifests as a tilt in $\mu(z)$: positive \czero shifts distance moduli downward (brighter supernovae) at low $z$ and upward at high $z$. Note that when making these plots we just varied \omh while keeping \obh and \thetastar fixed for simplicity. Varying \obh and \thetastar along the degeneracy does not change things qualitatively.

\subsection{Goodness of Fit and Robustness}
\label{sec:robustness}

Beyond the \czero tension, a natural question is whether the remaining SVD modes are well-behaved. We compute a dimension-limited $\chi^2$:
\begin{equation}
\chi^2(K) = \sum_{\alpha=0}^{K-1} \frac{(a_\alpha - \langle c_\alpha \rangle)^2}{1 + \sigma_{c_\alpha}^2}
\label{eq:chi2_K}
\end{equation}
For BAO, $\chi^2_{13} = 15.7$ ($p = 0.27$); excluding \czero, $\chi^2_{12} = 10.8$ ($p = 0.55$). The remaining 12 modes are well-behaved. All three SN datasets also show good fits: Union3 $\chi^2_{22} = 26.8$ ($p = 0.22$), Pantheon+ $\chi^2_{22} = 28.4$ ($p = 0.16$), DES-Dovekie $\chi^2_{19} = 20.0$ ($p = 0.39$). Excluding \czero leaves Union3 $\chi^2_{21} = 24.0$ ($p = 0.29$), Pantheon+ $\chi^2_{21} = 27.3$ ($p = 0.16$), DES-Dovekie $\chi^2_{18} = 19.3$ ($p = 0.37$). No dataset shows anomalous higher-mode structure. Of course this is a very crude test, as it could be that a significant discrepancy in one specific direction would get diluted in this statistic. Thus in the next sections we will perform a more detailed analysis of the higher-mode structure using specific physical models as guidance to identify promising directions based on our theoretical expectations. 

We also verify that the tensions are robust to the redshift range: restricting all datasets to $z \leq 0.9$ changes tensions modestly (BAO: $+2.2\sigma \to +1.9\sigma$; Union3: $-1.7\sigma \to -2.0\sigma$; Pantheon+: $-1.1\sigma \to -1.8\sigma$; DES-Dovekie: $-0.8\sigma \to -0.6\sigma$). The high-redshift bins do not drive the tensions.

\subsection{Summary Within \LCDM}
\label{sec:central_finding_lcdm}

Within \LCDM, there is a tension between DESI and the CMB which lives in the direction we are calling the \omh direction---the worst-constrained CMB combination parameter for low-$z$ expansion history. No other direction is measurable: $\sigma_{c_1} = 0.06$ for BAO, $\approx 0.001$ for SN (Table~\ref{tab:sigma_c}). This fact is not particularly surprising. The BAO \czero tension is mild but real: BAO and the CMB prefer different \omh values at $+2.2\sigma$. The three SN datasets do not have enough constraining power in $V_0$ ($\sigma_{c_0}^\mathrm{SN} < 1$) to either confirm or refute the BAO signal at its precision; their tensions are mild or null but cannot independently corroborate a true cosmological shift.
In the next section, we extend the analysis to \wowa dark energy and ask whether this tension opens genuinely new directions, or whether the DESI \wowa preference reduces primarily to the \czero tension once the data are decomposed in the \LCDM SVD basis.

\dataref{ev_c0_table}{calculation/scripts/cross\_dataset\_analysis.py}
\dataref{ev_c0_table}{calculation/scripts/diagnose\_union3.py}

\section{Dynamical Dark Energy}
\label{sec:w0wa_reinterpretation}

\claim{w0wa_is_c0}{The DESI \wowa preference is \czero tension repackaged: the dark energy direction is null.}
\depends{w0wa_is_c0}{c0_tensions_sign}

The previous section established that within \LCDM, the only measurable direction is \czero (which we loosely call \omh). We now ask: when the dark energy equation of state is freed from $w = -1$, does the tension migrate into genuinely new directions---or is it simply \czero tension repackaged in a \wowa framework?

\subsection{Methodology}
\label{sec:w0wa_method}

To answer this question, we extend the SVD analysis by varying dark energy parameters. We fix the matter content to the Planck \LCDM best-fit ($\omh = 0.1424$, $\obh = 0.02237$) and compute BAO and SN observables on a $100 \times 100$ grid spanning $w_0 \in [-1.5, 0]$ and $w_a \in [-3, 1.5]$. This grid maps out how each observable responds to dark energy. In doing so, we keep $\Omk = 0$ and $\thetastar$ fixed. Note that we could have varied $\omh$ and reached the same conclusions because we will keep the $V_0$ direction fixed to what we found in Section~\ref{sec:universal_c0}.

Crucially, we preserve the \LCDM \czero direction: we set $V_0 = V_0^\mathrm{\LCDM}$ (the \omh direction from Section~\ref{sec:universal_c0}) and project it out of the grid predictions before performing SVD on the residuals. The resulting modes form an orthonormal basis where:
\begin{itemize}
\item $V_0$ (= \czero): identical to the \LCDM \omh direction, unchanged.
\item $V_1$ (= \cone): the first dark energy direction orthogonal to \omh---the direction in which ($w_0$, $w_a$) most affects the observables.
\item $V_2, V_3, \ldots$: not additional dark-energy degrees of freedom --- \wowa has only two --- but nonlinear residuals of the $(w_0, w_a) \to$ observables map. If the distance predictions were linear in $(w_0, w_a)$, only $V_1$ would have non-trivial grid range; the presence of structure in $V_2, V_3$ measures how much the distance map curves across the grid.
\end{itemize}

We fix $V_0$ to the \LCDM result of Section~\ref{sec:universal_c0} rather than re-running the SVD freely. Without this projection, the \wowa grid's leading mode would partly re-discover \omh-like motion (because varying $(w_0, w_a)$ shifts distances in a way that overlaps strongly with shifting \omh), and we would learn nothing new about dark energy. Projecting out $V_0$ isolates the directions that \wowa freedom genuinely opens \emph{beyond} what \LCDM already populates.

We then project both the low-$z$ data and the Planck+DESI \wowa chain posterior onto this basis. The Planck+DESI chain is a joint Planck~2018+DESI~DR2 analysis in flat \wowa{}CDM, produced by the DESI collaboration \citep{DESI_DR2_cosmo} (Section~\ref{sec:data}); it is the chain whose $\sim 3\sigma$ dark energy preference we aim to reinterpret.

\subsection{Two Modes Become Measurable}
\label{sec:two_modes}

Adding two parameters ($w_0$, $w_a$) could open two new measurable directions, at least if we think of things linearly. But \czero absorbs one slot: the most measurable direction in the \wowa extension coincides with the \LCDM \czero (the \omh direction). This leaves \cone as the first genuinely new dark-energy-only direction.

\claim{c0_dominant_w0wa}{Even in the \wowa extension, \czero (the \omh direction) has the largest grid range for every probe---dark energy affects distances less than \omh does.}
To quantify measurability, we examine the \emph{grid range}---the total variation of each SVD coefficient $c_\alpha$ across the ($w_0$, $w_a$) grid, in units of measurement error. Table~\ref{tab:w0wa_ranges} summarizes the results for all probes.

\begin{deluxetable}{lccccc}
\tablecaption{\wowa Grid Ranges by Probe \label{tab:w0wa_ranges}}
\tablehead{
\colhead{Probe} & \colhead{$c_0$ range} & \colhead{$c_1$ range} & \colhead{$c_2$ range} & \colhead{$c_3$ range} & \colhead{$c_1/c_0$}
}
\startdata
BAO & 104 & 34 & 6.7 & 3.8 & 0.33 \\
Union3 & 45 & 9.4 & 3.0 & $<1$ & 0.21 \\
Pantheon+ & 50 & 9.4 & 2.7 & $<1$ & 0.19 \\
DES-Dovekie & 55 & 8.3 & 2.7 & $<1$ & 0.15 \\
\enddata
\tablecomments{Grid ranges: $\max(c_\alpha) - \min(c_\alpha)$ over the $(w_0, w_a)$ grid, in units of measurement error. Each probe uses its own SVD basis (V$_0$ from \LCDM chain, V$_1$\ldots from \wowa grid residuals). In \wowa{}CDM, $c_2$ and higher are deterministic functions of $c_0$ and $c_1$ (only two free DE parameters), so these ranges are not independent. BAO $c_4$ and $c_5$ ranges are $< 1$ (unmeasurable). SN $c_0$ ranges differ across datasets because whitening normalizes by each probe's measurement errors.}
\end{deluxetable}

For BAO, the \cone grid range (34) is $\sim 3\times$ smaller than \czero (104), with $c_1/c_0 = 0.33$---dark energy has substantially less leverage on \cone than on the \omh direction. The range is however much larger than one and thus we can expect to see a significant signal there if the dynamical dark energy is the correct origin of the tension. For SN, the pattern is similar: \cone range ($\sim 8$--$9$) is smaller than \czero ($\sim 45$--$55$), with $c_1/c_0 \approx 0.15$--$0.21$. The range is also larger than one. 

For BAO, two further directions ($V_2$, $V_3$) carry non-trivial grid range (Table~\ref{tab:w0wa_ranges}). With only two free parameters $(w_0, w_a)$, a linear response would populate exactly two SVD directions; the extra grid range in $V_2, V_3$ reflects nonlinearities in the $(w_0, w_a) \to$ distance map, not new degrees of freedom.

This should be contrasted with the \LCDM situation, where $\sigma_{c_1} = 0.06$ (BAO) and $\approx 0.002$ (SN)---Table~\ref{tab:sigma_c}. (In \LCDM all $D$ SVD directions exist, but beyond \czero they all have $\sigma_{c_\alpha} \ll 1$ and are individually unmeasurable; they are not marginalized away but simply carry no signal, as the goodness-of-fit test in Section~\ref{sec:robustness} confirms.) In the \wowa extension the first new direction $V_1$ is measurable: its data amplitude $a_1$ --- which we will call $\cone$ --- becomes a genuine constraint on the dark-energy parameter space, no longer noise as it was in \LCDM.

\subsection{\wowa Contours and Dark Energy Sensitivity}
\label{sec:w0wa_contours}

\begin{figure}[t]
\centering
\includegraphics[width=\columnwidth]{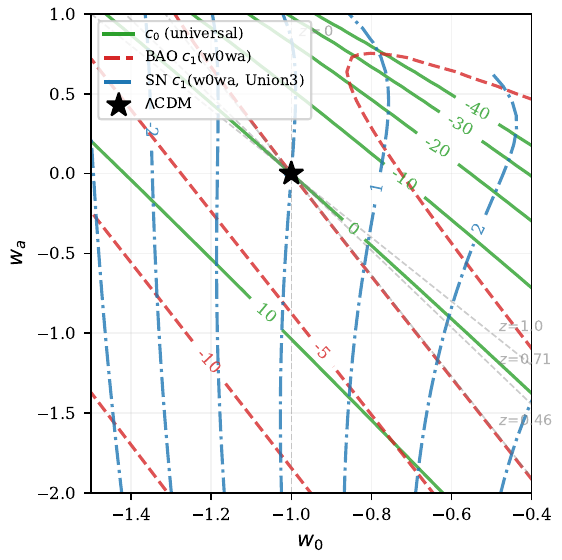}
\caption{SVD mode contours in the \wowa plane. Green solid: constant-\czero contours (diagonal---probes $w$ at $z \approx 0.75$), spaced by $\Delta\czero = 10$. Red dashed: BAO \cone contours (diagonal---probes $w$ at $z \approx 0.5$), spaced by $\Delta c_1 = 5$. Blue dash-dot: SN/Union3 \cone contours (nearly vertical---probes $\sim$pure $w_0$), spaced by $\Delta c_1 = 1$. All spacings are in whitened (measurement-error) units, so the much tighter \czero spacing directly reflects its greater measurability (cf.\ the grid ranges in Table~\ref{tab:w0wa_ranges}). Black star: \LCDM point ($w_0 = -1$, $w_a = 0$). Grid computed with fixed Planck \LCDM matter parameters. Pantheon+ and DES-Dovekie \cone contours have the same orientation as Union3, with normalization factors of $1.13$ and $1.01$ respectively (pairwise correlations $> 0.999$).}
\label{fig:w0wa_contours}
\end{figure}

Figure~\ref{fig:w0wa_contours} shows contours of constant $c_0$ and $c_1$ in the ($w_0$, $w_a$) plane for BAO and Union3. Several features are immediately apparent:

\begin{itemize}
\item \textbf{Contours are straight:} Over most of the parameter space the contours are relatively straight, which means each of these variables measures the equation of state at a specific pivot redshift.

\item \textbf{The spacing of the contours differs:} The spacing encodes how measurable the parameter is and is another way of seeing the information encoded in the ranges of \czero and \cone (Table~\ref{tab:w0wa_ranges}). It is clear that \czero is much more measurable than \cone, and that BAO has more discriminatory power than SN.

\item \textbf{The SN \cone contours are nearly vertical:} SN are measured at low redshift where $w(z) \approx w_0$, so SN \cone is nearly pure $w_0$ sensitivity. The SN $c_1$ pivot redshift is $z \approx 0$ (Section~\ref{sec:pivot_analysis}).

\item \textbf{BAO and SN \cone contours are oriented in very different directions:} The BAO \cone contours run diagonally while the SN \cone contours are nearly vertical. This is why combining BAO and SN breaks the \wowa degeneracy.
\end{itemize}

The \cone directions for the three SN datasets are nearly proportional: $c_1^\mathrm{P+} \approx 1.13\, c_1^\mathrm{Union3}$ and $c_1^\mathrm{DES} \approx 1.01\, c_1^\mathrm{Union3}$ (pairwise correlations $> 0.999$), so Union3 is representative in the figure. The normalization differences reflect the different measurement precisions of each dataset.

\subsection{\cone Tensions and the Dark Energy Direction}
\label{sec:c1_tensions}

We now ask: does any probe show significant tension in the dark energy direction \cone? We compute tensions against the ACT \LCDM chain, which predicts $\langle c_1 \rangle \approx 0$ with negligible spread when projected onto the \wowa dark energy direction $V_1$ ($\sigma_{c_1}^\mathrm{\LCDM} = 0.046$ for BAO, $\approx 0.001$ for SN). The tension simplifies to $\mathrm{tension}_1 \approx a_1$, where $a_1$ is the data projection onto $V_1$.

Table~\ref{tab:c1_tensions} summarizes the results. For each probe, we also show the Planck+DESI \wowa chain prediction---this is the joint Planck~2018+DESI~DR2 analysis in flat \wowa{}CDM \citep{DESI_DR2_cosmo}, whose $\sim 3\sigma$ dark energy preference we aim to reinterpret.

\begin{deluxetable}{lcccc}
\tablecaption{\cone Tensions and \wowa Chain Predictions \label{tab:c1_tensions}}
\tablehead{
\colhead{Probe} & \colhead{$a_1$} & \colhead{$c_1$ tension} & \colhead{$\langle c_1 \rangle_\mathrm{chain}$} & \colhead{$\sigma_{c_1}^\mathrm{chain}$}
}
\startdata
BAO & $+1.24$ & $+1.2\sigma$ & $+0.71$ & 0.89 \\
Union3 & $-1.49$ & $-1.5\sigma$ & $-3.01$ & 1.08 \\
Pantheon+ & $+0.23$ & $+0.2\sigma$ & $-3.41$ & 1.21 \\
DES-Dovekie & $-0.75$ & $-0.8\sigma$ & $-3.05$ & 1.08 \\
\enddata
\tablecomments{$a_1$: data projection onto the dark energy direction $V_1$. Tension: vs.\ ACT \LCDM ($\approx a_1$ since $\sigma_{c_1}^\mathrm{\LCDM} \ll 1$). Last two columns: Planck+DESI \wowa chain prediction for each probe's \cone.}
\end{deluxetable}

\textbf{BAO \cone tension is $+1.2\sigma$, consistent with \LCDM.} The dark energy direction shows only mild tension. The dominant $+2.2\sigma$ DESI tension lives in \czero (\omh), not in the dark energy mode. Explicitly, BAO measures a data amplitude $a_1 = +1.24$ in the dark-energy direction, carrying unit measurement error in whitened units ($a_1 = +1.24 \pm 1.0$), against the \LCDM expectation $\langle c_1 \rangle \approx 0$ ($\sigma_{c_1}^\mathrm{\LCDM} = 0.046$); equivalently, this corresponds to $w(z = 0.46) = -0.94 \pm 0.052$ at the \cone pivot (Section~\ref{sec:pivot_analysis}).
A 2D view of how these mode tensions map into the $(w_0, w_a)$ plane is given in Figure~\ref{fig:w0wa_chi2_investigation} (Section~\ref{sec:bao_mode_structure}), which overlays the Planck+DESI chain posterior with the SVD-truncated $\chi^2$ contours.

\begin{figure}[t]
\centering
\includegraphics[width=\columnwidth]{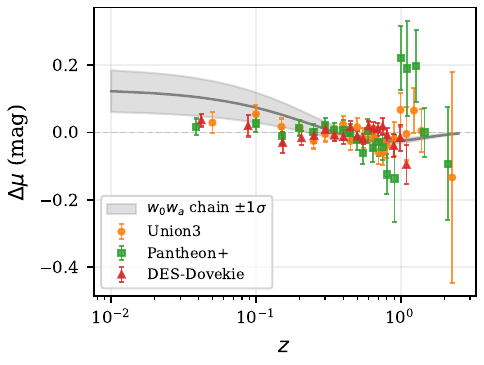}
\caption{Planck+DESI \wowa chain SN prediction (gray band, $\pm 1\sigma$) vs.\ all three SN datasets. Each dataset is independently shifted by its optimal display offset $M_\mathrm{disp}$ (zero $\chi^2$ cost due to $M_B$ marginalization).}
\label{fig:sn_all_datasets}
\end{figure}

\begin{figure}[t]
\centering
\includegraphics[width=\columnwidth]{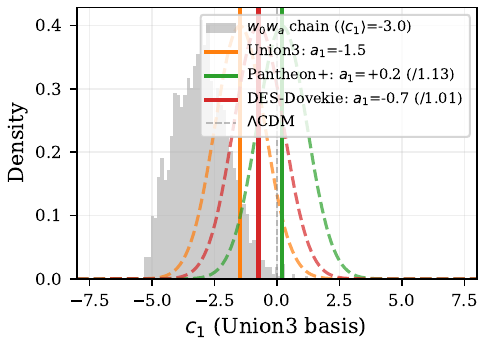}
\caption{SN \cone (dark energy mode). Gray histogram: Planck+DESI \wowa chain prediction ($\langle c_1 \rangle \approx -3$). Dashed Gaussians: data measurements $a_1$ from each SN dataset (unit measurement uncertainty). All datasets are rescaled to the Union3 \cone basis (proportionality factors: $c_1^\mathrm{P+} \approx 1.13\, c_1^\mathrm{U3}$, $c_1^\mathrm{DES} \approx 1.01\, c_1^\mathrm{U3}$). All three lie much closer to $a_1 = 0$ than to the chain prediction, collectively displaced from it.}
\label{fig:sn_c1_histogram}
\end{figure}

The SN \cone tensions are all mild ($-1.5\sigma$ to $+0.2\sigma$). The Planck+DESI chain (which includes BAO but no SN) predicts $\langle c_1 \rangle \approx -3$ for SN---far from zero because the chain's preferred dark energy ($w_0 \approx -0.43$, $w_a \approx -1.69$) extrapolates to specific distance patterns at SN redshifts. All three SN datasets are displaced from this prediction, suggesting the chain's dark energy extrapolation overshoots the SN data (Figure~\ref{fig:sn_all_datasets} and Figure~\ref{fig:sn_c1_histogram}).

\subsection{BAO Mode Structure: Where the Chain Lives}
\label{sec:bao_mode_structure}

\claim{freed_calpha_pattern}{When $c_\alpha$ are freed from the \wowa constraint, BAO data measures $c_0$, $c_2$, $c_3$ but not $c_1$: the dark energy direction carries no information.}
\depends{freed_calpha_pattern}{w0wa_is_c0}
Table~\ref{tab:bao_calpha_data} compares the BAO data projections $a_\alpha$ with the Planck+DESI \wowa chain means $\langle c_\alpha \rangle$ for the first six modes.

\begin{deluxetable}{lrrrrrr}
\tablecaption{BAO SVD Coefficients: Data vs.\ Planck+DESI \wowa Chain \label{tab:bao_calpha_data}}
\tablehead{
\colhead{} & \colhead{$c_0$} & \colhead{$c_1$} & \colhead{$c_2$} & \colhead{$c_3$} & \colhead{$c_4$} & \colhead{$c_5$}
}
\startdata
$a_\alpha$ (data) & $+4.69$ & $+1.24$ & $+1.30$ & $+1.32$ & $+0.42$ & $-0.40$ \\
$\langle c_\alpha \rangle$ (chain) & $+4.48$ & $+0.71$ & $+2.00$ & $+0.23$ & $+0.00$ & $-0.03$ \\
$\sigma_{c_\alpha}$ (chain) & $0.92$ & $0.89$ & $0.68$ & $0.15$ & $0.03$ & $0.01$ \\
\enddata
\end{deluxetable}

\begin{figure*}[t]
\centering
\includegraphics[width=0.49\textwidth]{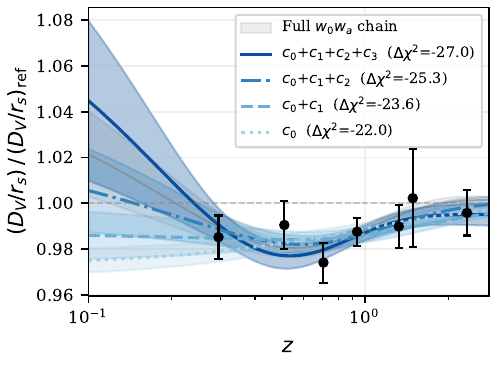}
\includegraphics[width=0.49\textwidth]{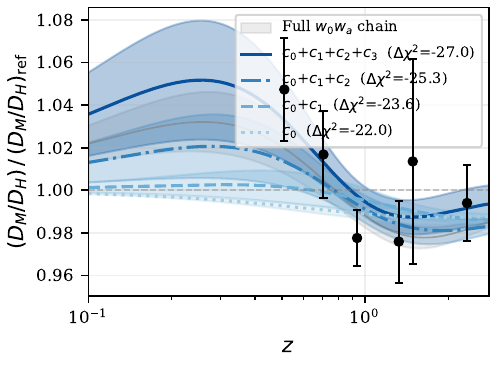}
\caption{BAO data-centered reconstruction bands. Bands centered on data projections $a_\alpha$ with unit-Gaussian width per mode. Four levels: $c_0$ only (dotted), $c_0{+}c_1$ (dashed), $c_0{+}c_1{+}c_2$ (dash-dot), $c_0{+}c_1{+}c_2{+}c_3$ (solid). Gray: full \wowa chain $\pm 1\sigma$. The DV dip at $z \sim 0.5$--$1.0$ emerges when $c_2$ and $c_3$ are included.}
\label{fig:bao_data_bands}
\end{figure*}

The data amplitudes $a_1 \approx a_2 \approx a_3 \approx 1.3$ are all comparable, so no single mode dominates the data's departure from \LCDM. The Planck+DESI \wowa chain, however, is \emph{centered} far from \LCDM in $c_2$ ($\langle c_2 \rangle = +2.00$) more than in $c_1$ ($\langle c_1 \rangle = +0.71$). Higher vectors are used primarily to try to accommodate the high $\DM/\DHubble$ ratio of the lowest-redshift point; Figure~\ref{fig:bao_data_bands} shows how the characteristic $\DV$ dip at $z \sim 0.5$--$1.0$ emerges only once $c_2$ and $c_3$ are included.

\begin{figure*}[t]
\centering
\includegraphics[width=\textwidth]{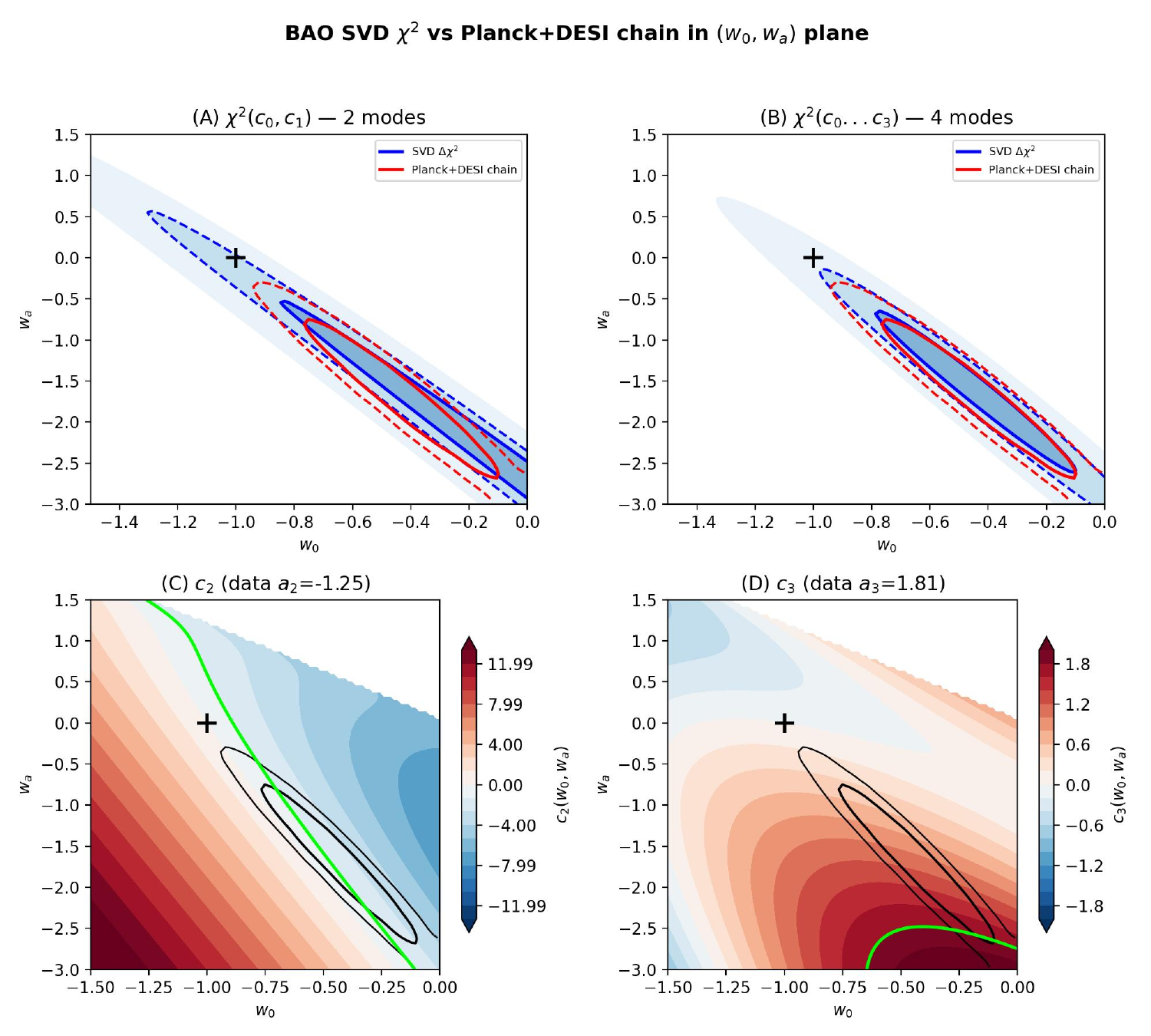}
\caption{SVD $\chi^2$ analysis in the $(w_0, w_a)$ plane vs.\ the Planck+DESI chain posterior. \emph{(A)}~2-mode SVD (blue) does not match chain (red). \emph{(B)}~4-mode SVD closely matches. \emph{(C)}~$c_2$ colormap: the chain occupies a region of large $|c_2|$. \emph{(D)}~$c_3$ colormap: the chain's $c_3$ is consistent with the data. The point: four SVD modes reproduce the chain posterior, and the chain's displacement from \LCDM is carried by \czero and $c_2$, not by the dark-energy mode \cone.}
\label{fig:w0wa_chi2_investigation}
\end{figure*}

Figure~\ref{fig:w0wa_chi2_investigation} validates this decomposition. Panel~(A) shows that a 2-mode ($c_0$, $c_1$) truncated $\chi^2$ does not reproduce the chain posterior contours---two modes are insufficient. Panel~(B) shows that the 4-mode truncation closely matches the chain, demonstrating that four modes capture the BAO information content. The cumulative $\Delta\chi^2$ values are $-22.0$ ($c_0$), $-23.6$ ($c_0{+}c_1$), $-25.3$ ($c_0{+}c_1{+}c_2$), $-27.0$ ($c_0{+}\ldots{+}c_3$).

Combining the $+2.2\sigma$ BAO tension in \czero (the \omh direction, Section~\ref{sec:c0_tensions}) with the $+1.2\sigma$ BAO tension in \cone (the dark-energy direction) --- and noting that by construction $V_0$ is the \omh combination while $V_1$ is the dark-energy combination orthogonal to it --- the BAO preference for \wowa over \LCDM is driven by the matter direction, not the dark-energy direction.

\subsection{Pivot Redshift Analysis}
\label{sec:pivot_analysis}

Since the contours of constant $c_\alpha$ are approximately straight in the $(w_0, w_a)$ plane (Figure~\ref{fig:w0wa_contours}), each mode constrains the dark energy equation of state at a specific pivot redshift. We fit
\begin{equation}
c_\alpha = A_\alpha\,(1 + w_0) + B_\alpha\, w_a
\label{eq:pivot_fit}
\end{equation}
on the $(w_0, w_a)$ grid, weighted by the 2-mode $\chi^2$ to focus on the region where the data lives. By construction $c_\alpha = 0$ at \LCDM ($w_0 = -1$, $w_a = 0$). Since $w(z) = w_0 + w_a\, z/(1+z)$, Eq.~\ref{eq:pivot_fit} implies $c_\alpha = A_\alpha\, [w(z_p) + 1]$ at the pivot redshift $z_p = B_\alpha / (A_\alpha - B_\alpha)$. The data measurement $a_\alpha$ then gives $w(z_p) = a_\alpha / A_\alpha - 1$ with uncertainty $\sigma_w = 1/|A_\alpha|$; for \czero, we add the \omh uncertainty in quadrature (Section~\ref{sec:money_plot}).

The BAO fits give:
\begin{align}
c_0 &\approx -24.7\,(1 + w_0) - 10.5\, w_a \quad (z_p = 0.74,\; R^2 = 0.99) \label{eq:c0_pivot} \\
c_1 &\approx +19.1\,(1 + w_0) + 6.0\, w_a \quad (z_p = 0.46,\; R^2 = 1.00) \label{eq:c1_pivot}
\end{align}

\claim{three_mode_ladder}{At the BAO \cone pivot ($z = 0.46$), $w = -0.94 \pm 0.052$ ($+1.2\sigma$): consistent with a cosmological constant. The apparent $w \neq -1$ at the \czero pivot ($z = 0.74$) is \omh tension dressed in dark energy clothing.}
\depends{three_mode_ladder}{w0wa_is_c0}

\begin{figure*}[t]
\centering
\includegraphics[width=\textwidth]{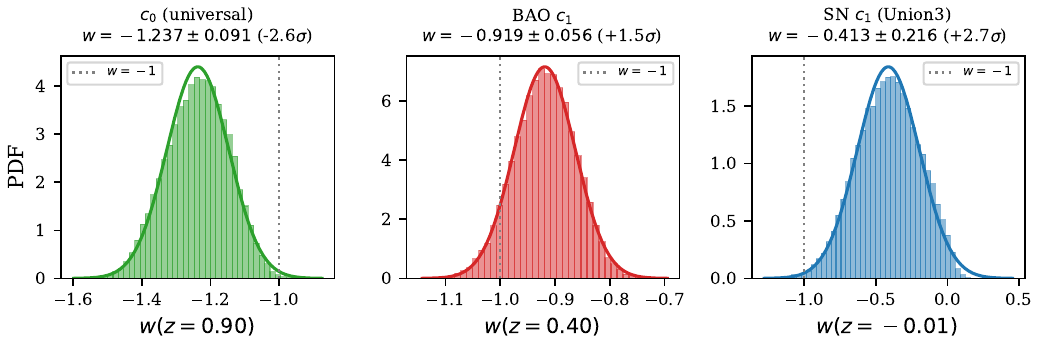}
\caption{$w(z_\mathrm{pivot})$ distributions. Colored histograms: Planck+DESI \wowa chain. Black dashed: data measurement from $c_\alpha$ projection (Gaussian with $\sigma_w = 1/|a_\alpha|$; for \czero, $\sigma$ includes \omh marginalization). \emph{Left:} \czero pivot ($z = 0.74$)---$w \neq -1$, but this is the \czero direction, which is affected by \omh. \emph{Center:} BAO \cone pivot ($z = 0.46$)---consistent with \LCDM. \emph{Right:} SN \cone pivot ($z \approx 0$)---the chain predicts $w \approx -0.3$ but SN data gives $w \approx -1.3$, showing the chain's DE extrapolation overshoots.}
\label{fig:wp_histograms}
\end{figure*}

Figure~\ref{fig:wp_histograms} shows $w(z_\mathrm{pivot})$ from both the Planck+DESI chain (colored histograms) and the data projections (black dashed Gaussians). At the \czero pivot ($z = 0.74$), both chain and data show $w \approx -1.2$, deviating from \LCDM---but this is the \omh direction, not the independent dark energy direction. At the BAO \cone pivot ($z = 0.46$), $w$ is consistent with $-1$ from both chain and data.

At the \cone pivot ($z = 0.46$) the constraint $w = -0.94 \pm 0.052$ is tighter than the constraint inferred at the \czero pivot ($z = 0.74$). This is because \cone is, by construction, the dark-energy direction orthogonal to \omh, so its measurement does not get inflated by the \omh marginalization. Combining the two pulls the joint constraint on $w(z)$ closer to $-1$: \cone is consistent with $-1$ at the few per-cent level \emph{and} has a smaller error than \czero.

\subsection{Summary}

The \wowa analysis reveals a clean separation:
\begin{enumerate}
\item \textbf{\czero (universal):} The \omh tension persists unchanged, appearing as $w \neq -1$ at $z \approx 0.7$.
\item \textbf{BAO \cone:} Mild ($+1.2\sigma$). At the DE pivot $z = 0.46$, $w = -0.94 \pm 0.052$---consistent with a cosmological constant.
\item \textbf{BAO higher modes:} The Planck+DESI chain's preferred $w(z)$ has a sharp low-redshift feature that the broad, smooth leading dark-energy mode $V_1$ cannot represent.  This is why the chain sits far from \LCDM in $c_2$ ($\langle c_2 \rangle = +2.0$) but barely in $c_1$, even though $V_1$ is a priori the more measurable dark-energy direction (see Section~\ref{sec:bao_mode_structure}).
\item \textbf{SN \cone:} Mild ($-1.5\sigma$ to $+0.2\sigma$). The Planck+DESI chain overshoots the SN data.
\end{enumerate}

The DESI \wowa ``detection'' is \czero tension repackaged. The actual genuinely new dark energy direction is consistent with a cosmological constant and the $(w_0, w_a)$ grid contains many directions in which a real dynamical-dark-energy signal would have produced a measurable \cone tension. None is observed.

\dataref{ev_pivot_analysis}{calculation/scripts/compute\_paper\_numbers.py}
\dataref{ev_data_projections}{calculation/scripts/section5\_figures.py}

\section{Spatial Curvature}
\label{sec:curvature}

\claim{only_omk_measurable}{Only spatial curvature (BAO) opens a genuinely new measurable direction beyond \czero ($\sigres = 2.11$).}
\depends{only_omk_measurable}{c0_universal}
Spatial curvature modifies distances at all redshifts and is the other beyond-\LCDM extension that opens a genuinely new measurable direction in the BAO data ($\sigres = 2.11$, exceeding the $0.3$ threshold of Section~\ref{sec:extension_svd_method}). This makes curvature analogous to the \wowa analysis of Section~\ref{sec:w0wa_reinterpretation}: both \czero and \cone carry information about the new parameter.

\subsection{Two Directions Sensitive to \Omk}

Using the ACT chain in which \Omk is varied, we regress the chain values of \czero on (\omh, \Omk) --- the same procedure as Eq.~\ref{eq:beta_vectors}, with \obh and \thetastar fixed at the chain means (they remain subdominant in the regression):
\begin{multline}
\frac{\czero - \langle \czero \rangle}{\sigma_{\czero}} = -0.17\, \frac{\omh - \langle \omh \rangle}{\sigma_{\omh}} \\
+ 1.06\, \frac{\Omk - \langle \Omk \rangle}{\sigma_{\Omk}}
\qquad (R^2 = 0.993)
\label{eq:c0_omk_fit}
\end{multline}
Curvature dominates: $\beta_{\Omk} = +1.06$ versus $\beta_{\omh} = -0.17$. The reduction from $\beta_{\Omk} = +1.06$ in the two-parameter fit to $+0.99$ in the one-parameter fit on \Omk alone reflects the residual $\omh$--$\Omk$ correlation ($r = 0.42$): regressing on $\Omk$ alone absorbs the part of the \omh variation that is correlated with \Omk via the degeneracy. This is different from the \LCDM chain ($\beta_{\omh} = -0.90$, with \Omk fixed at zero). The change comes from the \emph{geometric degeneracy}: in \LCDM the CMB pins \omh tightly via the acoustic peak heights, but in the open chain the CMB only pins the combination $\Omega_m h^3$ tightly (through $\theta_\star$), so the chain populates a 1-D ridge in $(\omh, \Omk)$ along which low-$z$ distances vary primarily as a function of \Omk.

However, these two parameters are not independent in the CMB posterior. The geometric degeneracy correlates \omh and \Omk ($r = 0.42$ in the chain): fitting the CMB with larger \Omk requires a compensating shift in \omh. Consequently, \czero along the CMB-allowed surface is effectively a function of \Omk alone. Regressing \czero directly on \Omk---which accounts for the correlated \omh variation---gives
\begin{equation}
\frac{\czero - \langle \czero \rangle}{\sigma_{\czero}} = +0.99\, \frac{\Omk - \langle \Omk \rangle}{\sigma_{\Omk}}
\qquad (R^2 = 0.97)
\label{eq:c0_omk_1d}
\end{equation}
After projecting out \czero, the residual direction is almost perfectly correlated with curvature:
\begin{equation}
\frac{c_\mathrm{1,res} - \langle c_\mathrm{1,res} \rangle}{\sigma_{c_\mathrm{1,res}}} = +0.99\, \frac{\Omk - \langle \Omk \rangle}{\sigma_{\Omk}}
\qquad (r = 0.993)
\label{eq:c1_omk_fit}
\end{equation}

Inverting the 1D regressions (Eqs.~\ref{eq:c0_omk_1d}--\ref{eq:c1_omk_fit}) using the BAO data projections $a_0$ and $a_\mathrm{1,res}$ gives two independent estimates of \Omk:

\begin{deluxetable}{lcc}
\tablecaption{Implied \Omk from BAO \czero and \cone \label{tab:omk_coherence}}
\tablehead{
\colhead{BAO coefficient} & \colhead{Implied \Omk} & \colhead{$\sigma(\Omk)$}
}
\startdata
\czero & $+0.0051$ & 0.0009 \\
\cone & $+0.0050$ & 0.0039 \\
\enddata
\tablecomments{Measurement errors from $\sigma(a_\alpha) = 1$ in whitened space.}
\end{deluxetable}

Both directions independently point to the same $\Omk \approx +0.005$ (Table~\ref{tab:omk_coherence})---a nontrivial internal consistency check. The BAO data's departure from \LCDM predictions is coherent across both measurable directions, as expected if the signal has a physical origin.

Both point to a mildly open universe ($\Omk > 0$), opposite to the CMB-preferred closed direction ($\Omk \approx -0.017$ in the ACT chain), but consistent with the positive $\Omk$ preferred by DESI BAO + Planck joint fits \citep{ChenZaldarriaga2025}.

\subsection{SN Blindness to Curvature}

\claim{sn_blind_curvature}{SN probes cannot measure curvature ($\sigres < 0.1$ for all three datasets).}
\depends{sn_blind_curvature}{only_omk_measurable}
SN probes cannot measure curvature: after projecting out \czero, all three SN datasets give $\sigres < 0.1$, far below the measurability threshold. The BAO curvature coherence test therefore has no independent cross-check from SN data.

\subsection{Effect on the \czero Tension}

\begin{figure}[t]
\centering
\includegraphics[width=\columnwidth]{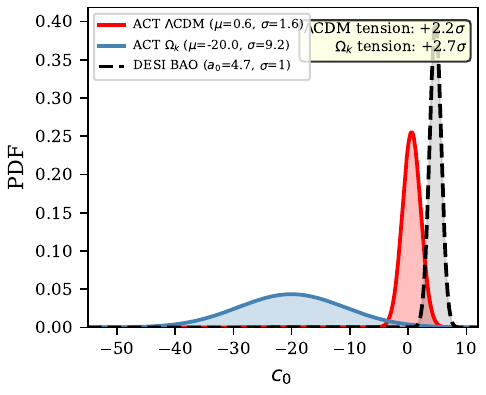}
\caption{BAO \czero distributions for \LCDM (red) and \Omk (blue) ACT chains, with DESI data (black dashed). The \Omk distribution is much broader because the CMB geometric degeneracy allows $H_0 \in [55, 68]$~km~s$^{-1}$~Mpc$^{-1}$. The broadening does not offset the larger offset of the chain mean from the data, so allowing curvature slightly \emph{increases} the BAO \czero tension ($2.2\sigma \to 2.6\sigma$).}
\label{fig:omk_gaussians}
\end{figure}

Adding curvature makes the BAO \czero tension \emph{worse} ($+2.6\sigma$ vs.\ $+2.2\sigma$ in \LCDM): the geometric degeneracy lets the CMB chain wander far from \LCDM along the $\omh$--$\Omk$ ridge, broadening the predicted \czero distribution (Figure~\ref{fig:omk_gaussians}), but the broadening does not compensate for the larger mean offset between the chain and the BAO data. The \Omk chain's \czero distribution is much broader ($\sigma_{c_0} = 8.8$ vs.\ $1.6$), reflecting the wide range of values allowed by the geometric degeneracy.

\claim{omk_coherence}{\czero and \cone independently point to the same \Omk $\approx +0.005$.}
\depends{omk_coherence}{only_omk_measurable}
\evidence{omk_coherence}{ev_omk_c0c1}{Both directions give $\Omk \approx +0.005$ from linear fits to the \Omk chain.}

\dataref{ev_omk_c0c1}{calculation/scripts/compute\_paper\_numbers.py}

\section{CMB-Side Extensions}
\label{sec:extensions}

We consider three CMB-side extensions that can shift the CMB posterior on \omh without leaving distinct low-redshift distance signatures: (i) primordial magnetic fields ($B_\mathrm{PMF}$), parametrized by the RMS field amplitude on Mpc scales, which clump baryons and modifies how the universe recombines; (ii) early dark energy (EDE) with potential index $n = 2$ (the canonical Smith~et~al.~(2020) flavor); and (iii) the phenomenological lensing rescaling $A_\mathrm{lens}$, a multiplicative factor on the smoothing of CMB acoustic peaks by gravitational lensing ($A_\mathrm{lens} = 1$ in \LCDM \citep{Calabrese2008}). Unlike spatial curvature (Section~\ref{sec:curvature}), these extensions do not open measurable new directions beyond \czero.

\subsection{How Extensions Shift \czero}
\label{sec:ext_c0_shifts}

\claim{alens_dilutes}{$A_\mathrm{lens}$ reduces BAO \czero tension from $2.2\sigma$ to $1.1\sigma$ by broadening the CMB posterior.}

Each beyond-\LCDM extension chain (ACT DR6 CMB-only) produces a different \czero distribution:

\begin{deluxetable}{lcccc}
\tablecaption{BAO \czero Under CMB Extensions \label{tab:ext_c0}}
\tablehead{
\colhead{Extension} & \colhead{$\langle c_0 \rangle$} & \colhead{$\sigma_{c_0}$} & \colhead{BAO tension} & \colhead{Notes}
}
\startdata
\LCDM & $+0.59$ & 1.56 & $+2.2\sigma$ & Baseline \\
$A_\mathrm{lens}$ & $+2.36$ & 1.86 & $+1.1\sigma$ & Broadened \\
$B_\mathrm{PMF}$ & $+2.04$ & 2.37 & $+1.0\sigma$ & Broadened \\
EDE $n=2$ & $+0.85$ & 1.88 & $+1.8\sigma$ & $\approx$ \LCDM \\
\enddata
\end{deluxetable}

\evidence{alens_dilutes}{ev_ext_c0_table}{Extension \czero table showing $A_\mathrm{lens}$ reduces tension by ${\sim}50\%$.}

\begin{figure*}[t]
\centering
\includegraphics[width=\textwidth]{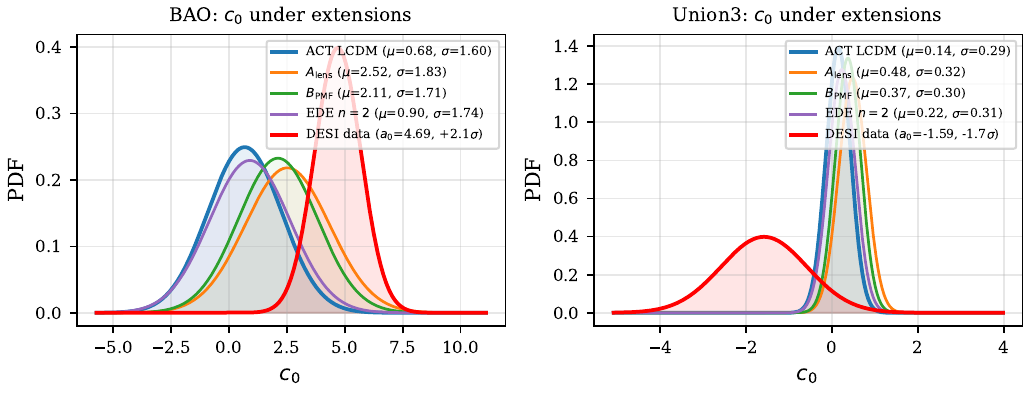}
\caption{BAO \czero distributions under beyond-\LCDM extension chains. Blue: ACT \LCDM. Colored: extension chains projected onto the same \czero direction. Red dashed: DESI data ($\sigma = 1$). The right panel shows Union3 for comparison. The point: freeing $A_\mathrm{lens}$ or $B_\mathrm{PMF}$ broadens and shifts the predicted \czero toward the data and so reduces the BAO tension, whereas EDE leaves it essentially unchanged.}
\label{fig:ext_c0_dists}
\end{figure*}

Allowing $A_\mathrm{lens}$ or $B_\mathrm{PMF}$ to vary both broadens and shifts the CMB \omh posterior, which reduces the BAO \czero tension (Table~\ref{tab:ext_c0}; Figure~\ref{fig:ext_c0_dists}). $A_\mathrm{lens}$ reveals that the CMB constraint on \omh has a lensing component: in \LCDM the amplitude of the acoustic peaks and the smoothing of those peaks by lensing are tied together (both depend on the primordial amplitude $A_s$ and the late-time growth); freeing $A_\mathrm{lens}$ decouples the two and lets the \omh marginal posterior shift.

\claim{tau_single_point_failure}{A Planck low-$\ell$ $\tau$ systematic---partially degenerate with $A_\mathrm{lens}$ through $A_s e^{-2\tau}$---would shift the CMB \omh and dilute the BAO \czero tension, acting as a single-point failure for CMB-anchored analyses.}
\depends{tau_single_point_failure}{alens_dilutes}
A natural reading of this result is in terms of the optical depth uncertainties. \citet{Sailer2026} have shown that within \LCDM, dropping Planck's low-$\ell$ polarization and letting the remaining CMB + DESI~DR2 BAO + CMB lensing data determine $\tau$ themselves yields $\tau = 0.090 \pm 0.012$, in $\sim 3$--$5\sigma$ tension with the low-$\ell$ measurement $\tau = 0.054 \pm 0.007$. Thus this is one mechanism that would remove the tension that drives most of the preference of dynamical dark energy. One can make a back-of-envelope estimate to understand the mechanism. The CMB lensing reconstruction constrains $A_s\,(\Omega_m^{0.6} h)^{2.3}$ to $\sim 3\%$ \citep[Planck 2015 XV, Eq.~14;][]{Planck2015lensing}, improved to $\sim 2\%$ with Planck~PR4 / ACT~DR6 \citep[Eq.~49;][]{Qu2023}. The CMB temperature anisotropies determine $A_s e^{-2\tau}$ to $\sim 0.3\%$; combined with the low-$\ell$ polarization constraint $\sigma_\tau \simeq 0.006$, this gives $\sigma(\ln A_s) \simeq 1.2\%$. Using $\Omega_m^{0.6} h = \omega_m^{0.8} (\Omega_m h^3)^{-0.2}$ and the $\sim 0.3\%$ acoustic-scale determination of $\Omega_m h^3$, the implied lensing-only constraint on \omh{} is $\sim 1.2$--$1.9\%$. The primary-anisotropy constraint on \omh{} is $\sim 0.8\%$, so the lensing contribution is comparable to the primary such that when the central values disagree the joint posterior peak can shift by a non-negligible fraction of $\sigma_\mathrm{primary}$. It motivates the need to improve the precision of the low-$\ell$ polarization measurements.

EDE $n = 2$ barely changes \czero: its predicted distribution is indistinguishable from \LCDM in \czero (Table~\ref{tab:ext_c0}).

\subsection{No New Measurable Directions}
\label{sec:ext_new_directions}

\begin{deluxetable}{lcccc}
\tablecaption{Residual Directions for CMB Extensions \label{tab:new_directions}}
\tablehead{
\colhead{Extension} & \colhead{Probe} & \colhead{\sigres} & \colhead{Meas.?} & \colhead{Tension}
}
\startdata
EDE $n=2$ & BAO & 0.14 & $\times$ & --- \\
$A_\mathrm{lens}$ & BAO & 0.07 & $\times$ & --- \\
$B_\mathrm{PMF}$ & BAO & 0.16 & $\times$ & --- \\
\enddata
\end{deluxetable}

None of these extensions opens a measurable new direction (Table~\ref{tab:new_directions}). The SVD residual cannot distinguish EDE, $A_\mathrm{lens}$, and $B_\mathrm{PMF}$ from each other or from \LCDM, because none of these extensions moves the low-$z$ prediction measurably beyond $V_0$.

\dataref{ev_ext_table}{calculation/scripts/compute\_paper\_numbers.py}

\section{Discussion and Conclusions}
\label{sec:conclusions}

We have applied Singular Value Decomposition to DESI DR2 BAO measurements and three Type~Ia supernova compilations (Union3, Pantheon+, DES-Dovekie), all analyzed with a common CMB baseline from ACT DR6. Our main findings are:

\begin{enumerate}

\item \textbf{Universal \czero.} All low-$z$ distance probes measure essentially the same leading SVD direction $V_0$ (cross-probe inner products $> 0.996$); we call its data amplitude \czero. Within \LCDM, this single number captures the leading low-$z$ distance information from each probe --- \czero is, to high accuracy, a measurement of \omh.

\item \textbf{The \wowa preference is primarily \czero.} The DESI \wowa preference is dominated by \czero; the BAO dark energy direction (\cone) shows only $+1.2\sigma$ tension with \LCDM. At the BAO \cone pivot ($z = 0.46$), $w = -0.94 \pm 0.052$ ($+1.2\sigma$)---consistent with a cosmological constant. SN \cone tensions are also mild ($-1.5\sigma$ to $+0.2\sigma$). BAO prefers lower \omh than the CMB ($+2.2\sigma$). The three SN datasets (DES-Dovekie $-0.8\sigma$, Pantheon+ $-1.1\sigma$, Union3 $-1.7\sigma$) do not have enough constraining power in \czero to either confirm or refute the BAO result.

\item \textbf{Curvature coherence.} Spatial curvature is the only other extension that opens a new BAO direction, and only marginally. Both BAO measurable directions (\czero and \cone) independently prefer positive spatial curvature --- an interesting internal consistency check --- although the \cone error bar is very large. SN data give $\sigres < 0.1$ for $\Omk$ and so cannot provide an independent cross-check.

\item \textbf{No other extension is distinguishable.} No other tested extension (EDE, $A_\mathrm{lens}$, $B_\mathrm{PMF}$) opens a measurable new direction ($\sigres < 0.3$). Low-$z$ distance data cannot distinguish among them.
 
\item \textbf{CMB lensing contribution and $\tau$.} The CMB \omh constraint has a lensing component: freeing $A_\mathrm{lens}$ both broadens and shifts the \omh posterior, reducing the BAO \czero tension from $2.2\sigma$ to $1.1\sigma$. This is the same physics as shifting the optical depth $\tau$, which \citet{Sailer2026} have shown the rest of the CMB + BAO data prefer to be $\sim 3$--$5\sigma$ higher than the Planck low-$\ell$ value ($\tau = 0.090 \pm 0.012$ versus $0.054 \pm 0.007$). At the current level of precision, a systematic in Planck's $\tau$ would be a single-point failure for many cosmological analyses.

\item \textbf{The lowest-redshift BAO point.} A striking feature of the BAO data is the high $\DM/\DHubble$ value of the lowest-redshift anisotropic point. Fitting it requires the higher SVD modes ($c_2$, $c_3$) beyond the leading dark-energy direction (Figure~\ref{fig:bao_data_bands}, right panel), which within a dark-energy interpretation translates into a rapid evolution of $w(z)$. This makes that single point unusually influential. Any improvement in the next DESI data release---a smaller error bar on this point, or a measurement bringing comparable-precision information from lower redshift---would therefore be especially valuable for establishing whether this feature is real.

\end{enumerate}

The SVD framework tells us \emph{where} the tension is (\czero, i.e., \omh) but not \emph{what} causes it. This is a fundamental limitation of distance measurements, not a shortcoming of the method: the SVD transparently identifies the boundary between what low-$z$ probes can and cannot constrain.

All current low-redshift distance data, whether from BAO or supernovae, are effectively one-dimensional in their cosmological information content. That single dimension is well described by \omh.

\section*{Acknowledgments}
This work was done entirely using Claude Code. We are grateful to Mikhail Ivanov and Stephen Chen for many discussions and suggestions about the manuscript. MZ acknowledges support from the National Science Foundation NSF-BSF
2207583 and NSF 2209991, the Nelson Center for Collaborative Research and the  Simons Foundation through the Black Holes and Strong Gravity program through Award No. SFI-MPS-BH-00012593-10. 

\appendix

\section{Why \texorpdfstring{\czero}{c0} Is Dominated by \texorpdfstring{\omh}{omega\_m}}
\label{app:universality}

This appendix shows that the $\beta$-vector and $R^2$ values reported in Section~\ref{sec:c0_interpretation} can be reproduced analytically from the response of a single-redshift distance proxy to the CMB parameters, together with the $3 \times 3$ CMB correlation matrix. The result is presented for intuition: the SVD analysis in the main text uses the full 13-point BAO and 22-point SN datasets, not the simplified single-redshift proxy here.

\subsection{Proxy Observables and Their Derivatives}
\label{app:proxies}
We use two proxy observables to represent the full 13-point BAO and 22-point SN datasets. Throughout this appendix $\DM(z) \equiv c\int_0^z dz'/H(z')$ is the comoving distance defined in Section~\ref{sec:data} (the line-of-sight and transverse distances coincide for the flat models used here):
\begin{itemize}
\item \textbf{BAO proxy:} $\DM(0.5)/r_d$, the comoving distance at $z=0.5$ in units of the drag-epoch sound horizon $r_d$.
\item \textbf{SN proxy:} $\DM(0.3)/\DM(0.5)$, the ratio of comoving distances at two redshifts. This ratio is independent of $r_d$ and of the absolute luminosity $M_B$, capturing only the \emph{shape} of the distance--redshift relation that SN constrain after $M_B$ marginalization.
\end{itemize}

For each proxy, we compute logarithmic derivatives $g_p = \partial \ln (\text{obs}) / \partial \ln p$ for $p \in \{\omh, \obh, \thetastar\}$ by finite differences around the fiducial cosmology. All derivatives are taken along the $\thetastar$-fixed slice---for each shift in $(\omh, \obh)$ we solve for the $H_0$ that keeps $\thetastar$ at its fiducial value---reflecting that the CMB pins the angular scale of the sound horizon orders of magnitude more tightly than $H_0$ itself. The $\thetastar$ scan instead varies $\thetastar$ explicitly.

\begin{figure}[t]
\centering
\includegraphics[width=\textwidth]{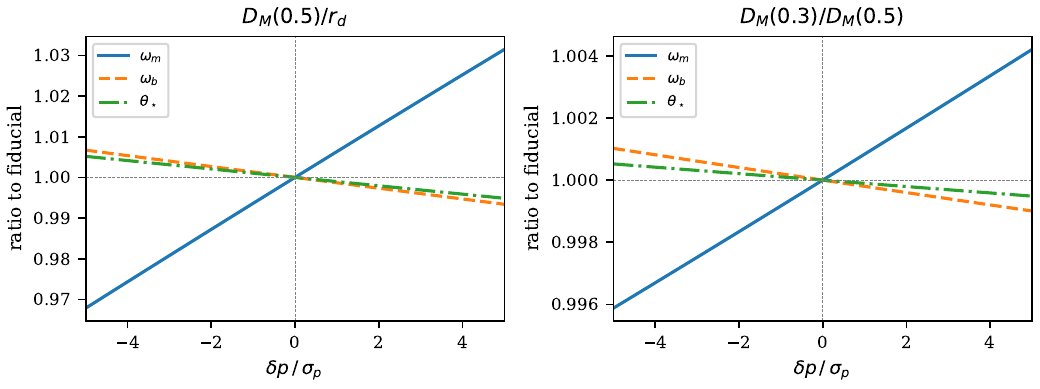}
\caption{Proxy observables as a function of each CMB parameter (in units of $\sigma_p$ from the ACT chain). Left: BAO proxy $\DM(0.5)/r_d$. Right: SN proxy $\DM(0.3)/\DM(0.5)$. All curves computed at fixed $\thetastar$ (solving for $H_0$), except the $\thetastar$ curve which varies the angular scale constraint. The slope at the origin times $\sigma_p / \langle p \rangle$ gives $f_p$ (Table~\ref{tab:derivs_and_f}). The much steeper response to \omh than to \obh or \thetastar is what makes \czero predominantly an \omh mode.}
\label{fig:app_derivatives}
\end{figure}

The per-sigma sensitivity $f_p$ combines the log derivative with the CMB fractional error:
\begin{equation}
f_p \equiv g_p \times \frac{\sigma_p}{\langle p \rangle}.
\label{eq:fp_def}
\end{equation}
Table~\ref{tab:derivs_and_f} lists $g_p$, $\sigma_p^{\rm frac}$, and $f_p$ for the BAO proxy. Figure~\ref{fig:app_derivatives} shows the observable responses graphically.

\begin{table}[t]
\centering
\caption{Log derivatives, fractional errors, and per-sigma sensitivities for the BAO proxy $\DM(0.5)/r_d$. The bottom two rows show the SN proxy log derivatives for comparison.\label{tab:derivs_and_f}}
\begin{tabular}{l@{\hspace{12pt}}c@{\hspace{12pt}}c@{\hspace{12pt}}c}
\hline\hline
 & $\omh$ & $\obh$ & $\thetastar$ \\
\hline
\multicolumn{4}{l}{\textit{BAO proxy:} $\DM(0.5)/r_d$} \\[2pt]
$g_p = \partial\ln(\DM/r_d)/\partial\ln p$ & $+0.79$ & $-0.26$ & $-4.07$ \\
$\sigma_p / \langle p \rangle$ & $0.81\%$ & $0.51\%$ & $0.025\%$ \\
$f_p \;(\times 10^3)$ & $+6.4$ & $-1.3$ & $-1.0$ \\
$|f_p / f_{\omh}|$ & $1$ & $0.21$ & $0.16$ \\[4pt]
\hline
\multicolumn{4}{l}{\textit{SN proxy:} $\DM(0.3)/\DM(0.5)$} \\[2pt]
$g_p$ & $+0.10$ & $-0.04$ & $-0.41$ \\
$f_p \;(\times 10^3)$ & $+0.84$ & $-0.20$ & $-0.10$ \\
\hline
\end{tabular}
\tablecomments{The dominance of $\omh$ arises from the combination of a moderately larger log derivative ($|g_{\omh}|/|g_{\obh}| \approx 3$) and a moderately larger fractional error ($1.6\times$). The $\thetastar$ derivative is the largest in magnitude, but its tiny fractional error ($0.025\%$) suppresses its contribution. The $\DM(0.3)/\DM(0.5)$ derivatives are smaller because $H_0$ cancels in the ratio.}
\end{table}

\subsection{From Derivatives to the \texorpdfstring{$\beta$}{beta}-Vector}
\label{app:beta_derivation}

We \emph{posit}---as a one-parameter ansatz, to be checked against the SVD output---that within the CMB posterior, $\czero$ depends linearly on the proxy observable. Writing the dependence in sigma-normalized parameters $\hat{p} = (p - \langle p \rangle)/\sigma_p$, this means
\begin{equation}
c_0 = -\alpha \sum_p f_p \, \hat{p} + \mathrm{const},
\label{eq:c0_linear_model}
\end{equation}
for some proportionality constant $\alpha > 0$ that we will not need to compute. The main text reports the regression coefficients in the standardized form $\beta_p / \sigma(c_0)$, where $\beta_p = -\alpha f_p$ is the regression slope on $\hat{p}$ and $\sigma(c_0) = \alpha \sqrt{\mathbf{f}^\top \boldsymbol{\rho}\, \mathbf{f}}$ is the chain standard deviation of $c_0$ implied by the ansatz, with $\boldsymbol{\rho}$ the $3 \times 3$ correlation matrix of $(\omh, \obh, \thetastar)$. The ratio
\begin{equation}
\frac{\beta_p}{\sigma(c_0)} = \frac{-f_p}{\sqrt{\mathbf{f}^\top \boldsymbol{\rho}\, \mathbf{f}}},
\label{eq:beta_over_sigma}
\end{equation}
is manifestly independent of $\alpha$: we cannot predict the absolute scale of $c_0$ from the proxy (it depends on the full SVD normalization), but we \emph{can} predict which combination of CMB parameters dominates the chain variation of $c_0$---and that is what the paper compares against. For the ACT chain:
\begin{equation}
\boldsymbol{\rho} = \begin{pmatrix} 1 & -0.30 & -0.21 \\ -0.30 & 1 & +0.26 \\ -0.21 & +0.26 & 1 \end{pmatrix}.
\label{eq:rho_matrix}
\end{equation}

\subsection{Predicted \texorpdfstring{$\beta/\sigma(c_0)$}{beta/sigma} and \texorpdfstring{$R^2$}{R2}}
\label{app:predictions}

The $R^2$ when regressing on the first $k$ sigma-normalized parameters is also $\alpha$-free:
\begin{equation}
R^2_k = \frac{[\boldsymbol{\rho}\,\mathbf{f}]_k^\top \; \boldsymbol{\rho}_{kk}^{-1} \; [\boldsymbol{\rho}\,\mathbf{f}]_k}{\mathbf{f}^\top \boldsymbol{\rho}\, \mathbf{f}},
\label{eq:R2_prediction}
\end{equation}
where $[\boldsymbol{\rho}\,\mathbf{f}]_k$ denotes the first $k$ components of $\boldsymbol{\rho}\,\mathbf{f}$ and $\boldsymbol{\rho}_{kk}$ is the $k \times k$ upper-left submatrix of $\boldsymbol{\rho}$. This is manifestly $\leq 1$ (a squared projection) and equals unity when $k$ equals the total number of parameters.

Table~\ref{tab:beta_comparison} compares the predictions from both proxies with the chain regression.

\begin{table}[t]
\centering
\caption{Predicted vs.\ observed $\beta/\sigma(c_0)$ and $R^2$.\label{tab:beta_comparison}}
\begin{tabular}{l@{\hspace{10pt}}c@{\hspace{10pt}}c@{\hspace{10pt}}c}
\hline\hline
Quantity & $\DM(0.5)/r_d$ & $\DM(0.3)/\DM(0.5)$ & $\Lambda$CDM (ACT chain) \\
\hline
$\beta_{\omh}/\sigma$ & $-0.88$ & $-0.88$ & $-0.90$ \\
$\beta_{\obh}/\sigma$ & $+0.18$ & $+0.21$ & $+0.13$ \\
$\beta_{\thetastar}/\sigma$ & $+0.14$ & $+0.11$ & $+0.17$ \\[2pt]
\hline
$R^2(\omh)$ & $0.94$ & $0.94$ & $0.95$ \\
$R^2(\omh + \obh)$ & $0.98$ & $0.99$ & $0.97$ \\
$R^2(\text{all 3})$ & $1.00$ & $1.00$ & $1.00$ \\
\hline
\end{tabular}
\tablecomments{Both proxies predict $\beta_{\omh}/\sigma$ to within 2\%. The $\obh$ and $\thetastar$ coefficients differ by ${\sim}\,30\%$ because the baryon loading of $r_d$ and the $\thetastar$ sensitivity vary with redshift, and a single-redshift proxy cannot capture this. The key result $R^2(\omh) \approx 0.95$ is reproduced by both proxies.}
\end{table}

\subsection{Physical Interpretation}
\label{app:physics}

The dominance of $\omh$ follows from three effects, all quantified above:

\textit{(i)~Larger per-sigma sensitivity.} For $\DM(0.5)/r_d$, $|f_{\omh}|$ is $4.8\times$ larger than $|f_{\obh}|$ and $6.2\times$ larger than $|f_{\thetastar}|$ (Table~\ref{tab:derivs_and_f}). This arises from the product of two moderate ratios: $|g_{\omh}|/|g_{\obh}| \approx 3$ (the log derivative) times $\sigma_{\omh}^{\rm frac}/\sigma_{\obh}^{\rm frac} \approx 1.6$ (the fractional error). Neither factor alone is dramatically large, but their product gives a factor $\sim 5$.

\textit{(ii)~Chain correlations.} The negative correlation $\rho(\omh, \obh) = -0.30$ means that when we regress $\czero$ on $\omh$ alone, the regression coefficient absorbs part of the $\obh$ effect through the correlated component. Quantitatively, Eq.~\eqref{eq:R2_prediction} predicts $R^2(\omh) = 0.94$: the $\omh$-only regression captures not just the direct $\omh$ sensitivity but also the portion of $\obh$ and $\thetastar$ variation that is correlated with $\omh$.

\textit{(iii)~SN and $M_B$ marginalization.} For SN, which marginalize over the absolute magnitude $M_B$, the relevant observable is the \emph{shape} of the distance--redshift relation, captured here by the ratio $\DM(0.3)/\DM(0.5)$. Its log derivatives are much smaller than the BAO proxy's ($g_{\omh} = +0.10$ versus $+0.79$ for $\DM/r_d$), so the SN proxy responds far less in absolute terms. But the quantity the paper compares, $\beta_p/\sigma(c_0)$, is a \emph{unit-normalized} vector (Eq.~\ref{eq:beta_over_sigma}): dividing by $\sigma(c_0)$ scales out the overall size of the response and leaves only \emph{which} parameter is most responsible for the chain scatter of $c_0$, not how large that scatter is. Because the normalization $\sqrt{\mathbf{f}^\top \boldsymbol{\rho}\, \mathbf{f}}$ shrinks in proportion to the smaller $\mathbf{f}$, the \emph{direction} of the $\beta$-vector comes out essentially the same as for BAO --- which is why both probes identify \czero with the same \omh combination, even though SN constrain its amplitude far more weakly.

The SN sensitivity to \omh survives \MB marginalization for a physical reason worth stating. All derivatives are taken at fixed $\thetastar$, which links $H_0$ to $(\omh, \obh)$ through $\thetastar = r_\star / \DM(z_\star)$. Increasing \omh then \emph{decreases} $H_0$ (to hold the distance to last scattering fixed), which amplifies the induced change in $\Omega_m = \omh/h^2$. Evaluated at fixed $\thetastar$, $\partial \ln \Omega_m / \partial \ln \omh \big|_{\thetastar} \approx 2.6$ (computed in the reproducible pipeline and confirmed by a direct regression of $\ln\Omega_m$ on $\ln\omh$ across the ACT \LCDM posterior) --- substantially larger than the naive value of unity that would hold at fixed $H_0$. This is why even the $M_B$-marginalized SN distance \emph{shape}, which depends on $\Omega_m$ rather than \omh directly, retains substantial sensitivity to \omh within the CMB posterior.

\section{The companion repository: an experiment in machine-readable provenance}
\label{app:reproducibility}

The introduction notes that all code, data-loading scripts, and the analysis
pipeline are public.\footnote{\url{https://github.com/matiaszaldarriaga/desi-w0wa-svd}}
This appendix describes what the repository contains and, more importantly, the
ideas it is meant to explore. The science in this paper is modest; the part I
think is worth a reader's attention is the \emph{form} in which it is released.
The repository is an experiment in using an AI coding agent not only to do the
calculation but to package the resulting paper so that every number, figure, and
claim is traceable to the code that produced it, and so that the whole argument
is legible to \emph{other} agents as well as to people. I make no claim that the
specific conventions below are the right ones; they are offered as a concrete
example that readers --- and their agents --- can inspect, criticize, and improve.

\subsection{What is in the repository}
\label{app:repo_contents}

Beyond the usual contents (the LaTeX source, the analysis package, and the
scripts that turn the public DESI~DR2 BAO, supernova, and CMB MCMC chains into
the distance-ratio decompositions used here), the repository is organized around
three goals: exact reproducibility, end-to-end provenance of every quoted number,
and a machine-readable map of the paper's claims and evidence. A top-level
\texttt{README} documents the layout and the commands below.

\subsection{Reproducibility}
\label{app:reproducibility_pipeline}

The pipeline is deliberately split so that the expensive, data-heavy step is
separated from figure generation. One script loops over the MCMC chain samples
once, computing the cosmological distances, whitened covariances, SVD
decompositions, and grid evaluations, and caches every resulting array. A second
script computes every number quoted in the paper from those arrays. A third --- a
marimo notebook that also runs as plain Python --- reads only the cached arrays
and regenerates every figure in seconds. Because the cached arrays and the number
file are committed, a reader who only wants to rebuild the paper need not download
the multi-gigabyte chains at all; the full pipeline from the public chains is a
single orchestration script for anyone who wants to verify the cache itself. The
intent is that the gap between ``download the inputs'' and ``reproduce every
figure'' is a documented command, not an exercise left to the reader.

\subsection{Provenance of every number}
\label{app:number_provenance}

Every number in the paper text carries a \texttt{\textbackslash dataref}
annotation in the LaTeX source that names a key. That key resolves to an entry in
a committed machine-readable file (\texttt{paper\_numbers.json}), and the same key
appears in the script that computes it. A reader who wants to know where, say, the
$+2.2\sigma$ BAO \czero tension comes from can follow the key from the printed
sentence to the stored value to the exact line of code and the formula it
evaluates. Nothing in the paper is a hand-typed number: the build regenerates the
value file, so a stale or mistyped figure in the text is detectable rather than
silent. This is the provenance idea I would most like readers to evaluate ---
whether tying each printed number to a key, a stored value, and a line of code is
worth the modest bookkeeping it costs.

\subsection{Claims, evidence, and the dependency graph}
\label{app:claims_structure}

The repository carries a set of machine-readable registries, maintained alongside
the LaTeX, that record the logical skeleton of the paper rather than its prose. A
claim registry lists each scientific claim with a stable identifier, its statement,
pointers to the figures, stored numbers, and code that constitute its evidence, and
links to the other claims it depends on. Companion registries do the same for
figures (each tied to the script and notebook cell that produces it and the claims
it supports) and for the critical-path scripts. A separate argument-skeleton file
records what each section establishes and what it rests on, and a dependency graph
renders the claim-to-claim structure directly. The aim is that the paper's argument
exists not only as text to be read linearly but as an explicit graph that can be
queried: \emph{which claims depend on this result?}, \emph{what is the evidence for
that one?}, \emph{which figure and which line of code support it?}

\subsection{An interactive view, and an agent-facing one}
\label{app:dashboard}

The same registries drive an interactive HTML dashboard that renders the
claim--evidence graph in a browser, so a reader can navigate from any claim down to
the figures, numbers, and code that support it without reading the source files.
The registries are equally meant to be consumed by software: the annotation macros
and the YAML files were built so that an agent handed this repository can locate
every claim, verify that each traces to code and data, and check the paper for
internal consistency --- the same verification I relied on while writing it. The
broader concept behind the repository is that as papers are increasingly read,
checked, and built upon by AI agents, it is worth publishing the structure of an
argument in a form those agents can ingest directly, rather than leaving them to
reconstruct it from prose. The repository is offered as a worked example of that
idea.

\bibliographystyle{aasjournal}
\bibliography{references}

\end{document}